\documentclass[a4paper]{article}
\usepackage{amsmath,amssymb,amsthm} 
\usepackage{graphicx,xcolor,ascmac,wrapfig} 
\newtheorem{teiri}{Theorem}
\newtheorem{katei}{Assumption}
\newtheorem{kei}{Corollary}

\newtheorem{hojo}{Lemma}

\def\ci{\perp\!\!\!\perp}
\usepackage[round]{natbib}
\usepackage{algorithm}
\usepackage{algpseudocode}
\usepackage{booktabs}
\usepackage{url}

\title{Bayesian ICA for Causal Discovery\thanks{Some preliminary results underlying this work were presented
at the 2024 IEEE International Symposium on Information Theory (ISIT)
and appeared in its proceedings \citep{suzuki2024generalization}.}}
\author{Joe Suzuki\\prof.joe.suzuki@gmail.com}

\begin{document}
\maketitle

\begin{abstract}
LiNGAM identifies causal orders by exploiting the non-Gaussianity and
mutual independence of structural disturbances. Several extensions
allow particular forms of latent confounding, but a general criterion
for comparing complete causal orders when the disturbances are
dependent remains lacking. We formulate ICA in a Bayesian manner and
propose Bayesian LiNGAM. 
For each candidate order, we use total correlation,
or multivariate mutual information, among the order-dependent
disturbances as a quantitative measure of confounding and estimate it  from data by Bayesian marginal
likelihoods. The causal order is selected by globally minimizing this
estimate. We prove the consistency of the estimator and implement the
global optimization using shortest-path search and dynamic
programming. In the no-confounding large-sample regime, the search
asymptotically requires the same number of score evaluations as
DirectLiNGAM. Bayesian LiNGAM therefore directly and globally
optimizes the original ICA dependence criterion, rather than relying
on greedy local decisions or a negentropy-based surrogate. Numerical
experiments show that it generally achieves better causal-order
recovery, particularly when latent confounding is present.

\end{abstract}

\section{Introduction}
\label{sec:introduction}

Causal discovery aims to infer causal relationships among variables from observational data. One of its central tasks is to determine a causal ordering of the variables, namely, an ordering in which no variable appearing later in the order can be a cause of a variable appearing earlier. Once such an ordering has been identified, the possible directions of the causal relations are substantially restricted, and the corresponding structural coefficients can be estimated by standard regression techniques.

In this study, we focus on the linear non-Gaussian acyclic model, commonly known as LiNGAM \citep{Shimizu2006}. Let $X_1,\ldots,X_p$ denote the observed variables. LiNGAM assumes that each variable is generated as a linear function of variables preceding it in the causal order, together with a disturbance term:
\[
X_i
=
\sum_{k(j)<k(i)} b_{ij}X_j+Z_i,
\]
where $k(i)$ denotes the position of $X_i$ in the causal ordering, $b_{ij}$ is a structural coefficient, and $Z_i$ is a non-Gaussian disturbance. The original LiNGAM formulation assumes that the disturbances $Z_1,\ldots,Z_p$ are mutually independent. This non-Gaussian independence makes it possible, under the model assumptions, to identify the causal ordering from observational data without specifying the ordering in advance.

The connection between LiNGAM and independent component analysis is straightforward. If the causal ordering and the structural coefficients are correct, removing the effects of the preceding variables from each variable produces the disturbance variables. These disturbances should be mutually independent when there is no latent confounding. Thus, causal ordering in LiNGAM can be regarded as a constrained ICA problem: among the linear transformations compatible with an acyclic structural equation model, one searches for a transformation whose resulting components are as independent as possible \citep{HyvarinenOja2000,Shimizu2006}.

Two standard approaches to this problem are ICA-LiNGAM and DirectLiNGAM. ICA-LiNGAM first applies an iterative ICA algorithm and subsequently resolves the permutation and scaling indeterminacies so that the estimated mixing structure is compatible with an acyclic model. Since general ICA algorithms rely on iterative numerical optimization, their finite-sample solutions can depend on initialization and optimization settings, and the algorithms may converge to local solutions. DirectLiNGAM was introduced to avoid this difficulty by identifying an exogenous variable and removing its effect successively \citep{shimizu11}. Pairwise likelihood-based measures provide another computationally simple implementation of this sequential principle \citep{HyvarinenSmith2013}. Under the exact LiNGAM assumptions and at the population level, these methods are theoretically well justified. In finite samples, however, DirectLiNGAM makes a sequence of locally evaluated choices, and a variable selected at an early step is not reconsidered at later steps. Thus, an early estimation error can influence all subsequent selections.

A further difficulty arises when latent confounding is present. A latent confounder is an unobserved variable that affects two or more observed variables. In an observed-variable structural equation model, such a common influence appears as statistical dependence among the disturbance terms. The mutual independence assumption of the standard ICA-LiNGAM and DirectLiNGAM models therefore corresponds to causal sufficiency, that is, the absence of unobserved common causes among the variables included in the analysis. Several extensions of LiNGAM have been developed to address latent variables and latent confounding, as reviewed in Section~\ref{sec:related-work}; see, for example, \citet{Tashiro2014}. The objective of the present study is not to explicitly identify the number or locations of latent confounders. Instead, we construct a complete causal ordering criterion that remains meaningful when the disturbances are not exactly independent.

The basic idea is simple. For every candidate ordering $\tau$, we regress each variable on the variables preceding it in that ordering and obtain an order-dependent disturbance vector
\[
Z^{\tau}
=
\left(
Z_1^{\tau},\ldots,Z_p^{\tau}
\right)^\top.
\]
If there is no latent confounding and $\tau$ is the correct causal ordering, these disturbances should be mutually independent. If the ordering is incorrect, some causal effects remain in the residuals, and dependence generally remains among them. When latent confounding is present, exact independence need not be achieved by any ordering. Nevertheless, different orderings can produce different degrees of residual dependence. This motivates selecting the ordering for which the remaining dependence is smallest.

To quantify this dependence, we use multivariate mutual information in the form of total correlation \citep{watanabe1960information}. For a candidate ordering $\tau$, it is defined by
\[
K_\tau
=
D_{\mathrm{KL}}
\left(
p_{Z^\tau}
\mathrel{\Big\|}
\prod_{j=1}^{p}p_{Z_j^{\tau}}
\right)
=
\sum_{j=1}^{p}h\left(Z_j^{\tau}\right)
-
h\left(Z^{\tau}\right),
\]
where $D_{\mathrm{KL}}$ denotes the Kullback--Leibler divergence and $h(\cdot)$ denotes differential entropy. This quantity is nonnegative and equals zero if and only if the disturbances are mutually independent. In the present structural equation model, latent confounding is represented by dependence among the disturbances. We therefore use $K_\tau$ as an operational measure of the confounding-induced dependence remaining under the ordering $\tau$, and define the target ordering by
\[
\tau^\ast
=
\arg\min_{\tau}K_\tau.
\]
In the absence of confounding, the minimum is ideally zero and is attained at the true causal ordering. In the presence of confounding, the criterion selects the ordering that makes the estimated disturbances as close to mutually independent as possible.

We formulate this independence criterion in a Bayesian manner. More specifically, we introduce a Gaussian copula model with non-Gaussian Student-$t$ marginal distributions and construct entropy and multivariate mutual information estimates from Bayesian marginal likelihoods. The marginal likelihood integrates the likelihood over the model parameters rather than evaluating it only at a single fitted parameter value. It therefore provides a principled Bayesian score for comparing the disturbance distributions produced by different candidate orderings. We show that, under the conditions stated in Section~3, the proposed Bayesian estimate $\widehat{K}_\tau$ consistently estimates its population counterpart:
\[
\widehat{K}_\tau
\xrightarrow{\mathrm{p}}
K_\tau
\qquad
(n\to\infty).
\]
Thus, the quantity minimized by the proposed algorithm has a direct information-theoretic interpretation, and its Bayesian estimate is asymptotically justified.

An important feature of the proposed formulation is that it does not require direct estimation of the full high-dimensional multivariate mutual information for every candidate ordering. The transformation from the observed variables to the order-dependent disturbances is triangular with unit diagonal and therefore has determinant one. Consequently,
\[
h\left(Z^\tau\right)
=
h({X})
\]
for every candidate ordering $\tau$. The joint entropy term is common to all orderings and does not affect their comparison. Hence, minimizing $K_\tau$ is equivalent to minimizing
\[
\sum_{j=1}^{p}h\left(Z_j^\tau \right).
\]
This reduction allows the global criterion to be evaluated using accurately estimated low-dimensional, and ultimately univariate, marginal-likelihood terms, without directly estimating a $p$-dimensional mutual information quantity.

Finally, we optimize this criterion globally over the possible variable orderings. Each partial ordering is represented by a subset of variables, and adding one variable to the end of a partial ordering corresponds to an edge in a directed subset graph. A complete causal ordering is therefore represented by a path from the empty set to the set of all variables. The ordering minimizing the proposed Bayesian ICA criterion can then be obtained as a shortest path, with dynamic programming used to reuse the scores of common partial orderings. In contrast to an iterative ICA solution or a sequence of irrevocable local choices, this construction finds a globally optimal ordering with respect to the proposed estimated criterion.

The numerical experiments compare the proposed method with ICA-LiNGAM and DirectLiNGAM under both unconfounded and confounded settings. When no confounding is present, the existing methods already attain performance close to the attainable ceiling, and the proposed method generally performs comparably and sometimes improves upon them. When confounding is present, the benefit of globally minimizing the estimated residual dependence becomes more visible, and the proposed method frequently achieves higher causal-order recovery rates and smaller pairwise ordering errors. These findings suggest that a Bayesian ICA criterion combined with global optimization provides a useful alternative to the standard LiNGAM procedures, particularly when the disturbance independence assumption is only approximately satisfied.

\subsection{Related Work}
\label{sec:related-work}

LiNGAM was introduced by \citet{shimizu06} as an identifiable
linear structural equation model with mutually independent non-Gaussian
disturbances. Standard estimation procedures include ICA-LiNGAM,
DirectLiNGAM \citep{shimizu11}, and the pairwise likelihood-ratio
approach of \citet{HyvarinenSmith2013}.

Extensions of LiNGAM have been developed for structural vector
autoregressive models \citep{hyvarinen2010svar}, linear systems containing
cycles \citep{lacerda2008cyclic}, groups of variables
\citep{kawahara2010grouplingam}, multiple data sets sharing a common
causal ordering \citep{shimizu2012joint}, longitudinal observations
\citep{kadowaki2013longitudinal}, and functional observations
\citep{yang2022functional}. Related identifiable causal models include
nonlinear additive-noise and post-nonlinear models
\citep{hoyer2008nonlinear,zhang2009postnonlinear,uemura2022postnonlinear},
discrete additive-noise models and BExSAM
\citep{peters2011discrete,inazumi2011bexsam}, and linear causal models
containing both continuous and discrete variables \citep{zeng2022mixed}.

Several methods address latent variables or latent confounding.
\citet{hoyer2008hidden} formulated latent-variable LiNGAM as an
overcomplete ICA problem. Other approaches include LiNGAM for latent
factors \citep{shimizu2009latent}, Pairwise LvLiNGAM
\citep{entner2011unconfounded}, models with latent Gaussian confounders
\citep{chen2013latentgaussian}, ParceLiNGAM \citep{Tashiro2014}, and
Bayesian bivariate causal-direction estimation with individual-specific
confounders \citep{shimizu2014bayesian}. Likelihood-Free Overcomplete ICA
was proposed by \citet{ding2019lfoica}, and
\citet{salehkaleybar2020latent} studied identification of linear
non-Gaussian causal models with latent variables. Further methods include
Multi-Domain LiNGAM \citep{zeng2021multidomain}, RCD
\citep{maeda2020rcd}, MLC-LiNGAM \citep{chen2022multiple}, and FRITL
\citep{chen2021fritl}. \citet{WangDrton2023} established
identification results for bow-free acyclic path diagrams with dependent
non-Gaussian errors, while \citet{cai2023cumulants} developed a
higher-order-cumulant approach under specific latent-component
conditions.

Mutual information is a classical dependence criterion for ICA.
\citet{comon1994ica} formulated ICA as the minimization of the
Kullback--Leibler divergence between the joint distribution of the
transformed components and the product of their marginal distributions.
Methods that estimate entropy or mutual information directly include
those of \citet{pham2002mutual}, RADICAL
\citep{learnedmiller2003ica}, MILCA \citep{stogbauer2004least}, and
MISEP \citep{almeida2003misep}.

Bayesian ICA has also been developed using nonparametric and hierarchical
source models. \citet{shen2016rateadaptive} placed priors on the unknown
block structure, mixing matrix, and source densities and established
rate-adaptive posterior contraction. More recently,
\citet{datta2025bayesianica} proposed a hierarchical Bayesian ICA model
with super-Gaussian source priors and developed EM and MCMC procedures,
together with posterior contraction and local asymptotic normality
results for the unmixing matrix.

Shortest-path formulations have been used for confounding-aware
causal-order estimation. \citet{suzuki2022binary} developed such a
criterion for binary variables. For linear continuous models,
LiNGAM-MMI measures dependence among the order-dependent disturbances
by total correlation and obtains a globally minimizing order through
shortest-path search \citep{suzuki2024generalization}.
\citet{ong2024redefining} subsequently studied pairwise likelihood-ratio
edge scores, prior-order information, and path enumeration within this
framework.


\subsection{Contributions}
\label{sec:contributions}

The main contributions of this study are summarized as follows.

\begin{enumerate}
\item
\textbf{An information-theoretic formulation of causal ordering under dependent disturbances.}
We formulate causal-order estimation as the minimization of the multivariate mutual information, or total correlation, among order-dependent disturbances. This gives a quantitative measure of the dependence remaining among the disturbances and provides a well-defined target even when latent confounding prevents exact independence.

\item
\textbf{A Gaussian-copula Bayesian formulation of the ICA independence criterion and its consistency.}
We propose a Gaussian copula model with non-Gaussian Student-$t$ marginals for the order-dependent disturbances and construct a Bayesian estimator of total correlation from marginal likelihoods. We prove that the resulting estimator converges in probability to the corresponding population multivariate mutual information under the stated conditions.

\item
\textbf{Avoidance of direct high-dimensional mutual information estimation.}
By exploiting the unit-determinant triangular transformation associated with an acyclic structural equation model, we show that the joint entropy term is invariant across candidate orderings. The global ordering criterion can therefore be compared through a sum of lower-dimensional marginal entropy terms, avoiding direct estimation of high-dimensional multivariate mutual information.

\item
\textbf{Global optimization by shortest-path search and its theoretical analysis.}
We formulate global causal-order search as a shortest-path problem on a subset graph and obtain the globally optimal ordering with respect to the proposed Bayesian score through shortest-path search and dynamic programming. Scores for common partial orderings are shared, so the method does not necessarily enumerate all $p!$ orderings. We prove that, in the no-confounding large-sample regime, the search asymptotically requires the same number of score evaluations as DirectLiNGAM. Numerical experiments further corroborate this theoretical result and show that the search examines only a limited part of the state space, with a practical computational burden comparable to that of the standard methods in the dimensions considered.

\item
\textbf{Empirical comparison with established LiNGAM methods.}
Through numerical experiments with and without latent confounding, we demonstrate that the proposed method generally performs comparably to or better than ICA-LiNGAM and DirectLiNGAM. The improvement is particularly evident under confounding, while comparable performance is retained in the unconfounded settings in which the existing methods already perform close to optimally.
\end{enumerate}

The remainder of this paper is organized as follows. Section~2 reviews the LiNGAM model, ICA-LiNGAM, DirectLiNGAM, and the Gaussian copula and Bayesian marginal-likelihood tools used in this study. Section~3 proposes the Gaussian-copula Bayesian ICA criterion, derives its lower-dimensional representation, and establishes its theoretical properties, including the consistency of the multivariate mutual information estimator. Section~4 formulates causal-order search as a shortest-path problem, develops the shortest-path and dynamic-programming algorithms, establishes the asymptotic result on the number of score evaluations, and presents the numerical experiments comparing the proposed method with the existing LiNGAM procedures. Section~5 concludes the paper and discusses directions for future research.

\section{Preliminaries}
This section reviews the background needed to understand the proposed
method. Sections~2.1 and~2.2 describe DirectLiNGAM and ICA-LiNGAM,
respectively, while Section~2.3 introduces the Gaussian-copula model
and its marginal likelihood used in our Bayesian formulation.

\subsection{DirectLiNGAM}

Given observed samples $x^n=(x_1,\ldots,x_n)$ and $y^n=(y_1,\ldots,y_n)\in\mathbb{R}^n$ ($n\geq 1$) of random variables $X$ and $Y$, our goal is to identify which of the two variables is the cause and which is the effect.
In the following discussion, we say that $X$ and $Y$ are respectively the cause and the effect if there exist $0\not=a\in\mathbb{R}$ and a random variable $Z$ such that
\begin{equation}\label{eq1}
Y = aX + Z,\quad X \ci Z,
\end{equation}
and conversely, we say that $Y$ and $X$ are respectively the cause and the effect if there exist $0\not=a'\in\mathbb{R}$ and a random variable $Z'$ such that\footnote{$U \ci V$ means that the random variables $U$ and $V$ are independent.}
\begin{equation}\label{eq2}
X = a'Y + Z',\quad Y \ci Z'.
\end{equation}
This is known as the additive noise model \citep{kano}.

Although details are omitted here \citep{shimizu11}, under either
model~(\ref{eq1}) or~(\ref{eq2}), if $(X,Y)$ is jointly Gaussian,
then the reverse model also holds, making causal identification
impossible.
This motivates the non-Gaussian assumption in LiNGAM
(Linear Non-Gaussian Acyclic Model) \citep{shimizu06}.
DirectLiNGAM \citep{shimizu11} directly estimates the causal order
under this model.
Concretely, after centering $x^n$ and $y^n$, define
\[
v(x^n)=\frac{1}{n}\sum_{i=1}^n x_i^2,\qquad
v(y^n)=\frac{1}{n}\sum_{i=1}^n y_i^2,\qquad
c(x^n,y^n)=\frac{1}{n}\sum_{i=1}^n x_i y_i.
\]
The residuals are then computed as
\[
y_x^n := y^n - \frac{c(x^n,y^n)}{v(x^n)}x^n,\qquad
x_y^n := x^n - \frac{c(x^n,y^n)}{v(y^n)}y^n.
\]
From the samples, one then assesses which of the two pairs,
$(x^n,y_x^n)$ or $(x_y^n,y^n)$, is closer to independence.

Earlier implementations of DirectLiNGAM used kernel-based independence
tests such as HSIC (Hilbert--Schmidt Independence Criterion; \citealp{hsic}).
In the current Python implementation \citep{IkeuchiEtAl2023}, however, the pairwise
likelihood-ratio criterion PW-LiNG \citep{HyvarinenSmith2013} is used as the default measure.
For a standardized random variable $U$, its non-Gaussianity is measured
by the negentropy defined as
\begin{equation}\label{eq:direct-negentropy}
J(U)
=
h(U_{\mathrm{gauss}})-h(U) \geq 0    
\end{equation}
with equality if and only if $U$ is Gaussian,
where $U_{\mathrm{gauss}}$ is a Gaussian random variable with the same
variance as $U$.
Note that the Gaussian distribution maximizes differential
entropy $h(\cdot)$ among all distributions with a fixed variance. 
PW-LiNG uses a
univariate negentropy approximation to construct the score \citep{HyvarinenOja2000}
\[
\widehat{D}_{XY}
:=
\left\{
\widehat{h}(y^n)
+
\widehat{h}(x_y^n)
\right\}
-
\left\{
\widehat{h}(x^n)
+
\widehat{h}(y_x^n)
\right\}
=
\widehat{J}(x^n)
+
\widehat{J}(y_x^n)
-
\widehat{J}(y^n)
-
\widehat{J}(x_y^n),
\]
where 
all variables and residuals are standardized before their
entropies are evaluated. 
For a standardized sample $u^n=(u_1,\ldots,u_n)$, $\widehat{J}(u^n)$ and $\widehat{h}(u^n)$ are defined by
\[
\widehat{J}(u^n)
=
79.047
\left\{
\frac{1}{n}\sum_{i=1}^{n}\log\cosh(u_i)-0.37457
\right\}^{2}
+
7.4129
\left\{
\frac{1}{n}\sum_{i=1}^{n}
u_i\exp\left(-\frac{u_i^{2}}{2}\right)
\right\}^{2},
\]
$$
\widehat{h}(u^n)
=
\frac{1}{2}\log(2\pi e)
-
\widehat{J}(u^n).$$
Thus, unlike direct estimators of bivariate mutual information
\citep{Kraskov,mine}, PW-LiNG compares the two causal directions using
univariate negentropy estimates.
PW-LiNG is computationally simple and has been reported to exhibit good
empirical performance, particularly for relatively small samples and
noisy data \citep{HyvarinenSmith2013}.
If $X$ causes $Y$, then $X\ci Z$ and
$\widehat{D}_{XY}$ is expected to be nonnegative, whereas its sign is
expected to be reversed under the direction $Y\to X$.

\begin{figure}
{\setlength{\unitlength}{0.65mm}
\small
\begin{center}
\begin{picture}(200,55)(0,0)

\put(10,5){\circle{10}}
\put(10,5){\makebox(0,0){$X$}}
\put(50,5){\circle{10}}
\put(50,5){\makebox(0,0){$Y$}}
\put(15,5){\vector(1,0){30}}
\put(45,30){\dashbox{0.1}(10,10){$Z$}}
\put(75,30){\dashbox{0.1}(10,10){$Z$}}

\put(80,5){\circle{10}}
\put(80,5){\makebox(0,0){$X$}}
\put(120,5){\circle{10}}
\put(120,5){\makebox(0,0){$Y$}}
\put(80,35){\makebox(0,0){$Z$}}
\put(115,5){\vector(-1,0){30}}

\put(150,5){\circle{10}}
\put(150,5){\makebox(0,0){$X$}}
\put(190,5){\circle{10}}
\put(190,5){\makebox(0,0){$Y$}}

\put(165,30){\dashbox{0.1}(10,10){$Z$}}
\put(155,5){\dashbox{1}(30,0)}

\put(0,20){(a)}
\put(90,20){(b)}
\put(140,20){(c)}

\put(20,20){\makebox(0,0){$X\rightarrow Y$}}
\put(110,20){\makebox(0,0){$X\leftarrow Y$}}

\put(20,35){\makebox(0,0){\color{blue}no confounding}}
\put(110,35){\makebox(0,0){\color{blue}no confounding}}

\put(170,13){\makebox(0,0){\color{red}confounding}}
\put(171,8){\makebox(0,0){\color{red}present}}

\thicklines
\put(80,30){\color{blue}\vector(0,-1){20}}
\put(50,30){\color{blue}\vector(0,-1){20}}
\put(167,31){\color{red}\vector(-2,-3){14}}
\put(173,31){\color{red}\vector(2,-3){14}}
\end{picture}
\end{center}
}
\caption{When noise affects only a single variable (shown in blue), LiNGAM can identify the causal direction as either $X\rightarrow Y$ or $Y\rightarrow X$, provided that at least one of $X$ or $Y$ is non-Gaussian.  
When noise affects multiple variables (shown in red), the causal direction cannot generally be identified by the standard
bivariate LiNGAM criterion
\label{fig1}}
\end{figure}

LiNGAM assumes from the outset that the statistical model is either~(\ref{eq1}) or~(\ref{eq2}).
However, in real data, the conditions $X\ci Z$ or $Y\ci Z'$
(Fig.~\ref{fig1}(a)(b)) rarely hold exactly, and confounding is common,
as illustrated in Fig.~\ref{fig1}(c). In such cases, PW-LiNG provides a
relative preference between the two directions rather than establishing
exact independence.

To illustrate the multivariate case, let
$z^n=(z_1,\ldots,z_n)\in\mathbb{R}^n$, and define
$z_x^n$, $z_y^n$, $x_z^n$, and $y_z^n$ analogously to
$y_x^n$ and $x_y^n$. The second-stage residuals include
\[
y_{xz}^n := y_x^n - \frac{c(y_x^n, z_x^n)}{v(z_x^n)}z_x^n,\qquad
z_{xy}^n := z_x^n - \frac{c(y_x^n, z_x^n)}{v(y_x^n)}y_x^n.
\]
The remaining second-stage residuals are defined analogously.
The causal order is determined according to
\[
\left\{
\begin{array}{lll}
x^n \ci \{y_x^n, z_x^n\} & \Longrightarrow &
\left\{
\begin{array}{lll}
y_x^n \ci z_{xy}^n & \Longrightarrow & X \rightarrow Y \rightarrow Z,\\
z_x^n \ci y_{xz}^n & \Longrightarrow & X \rightarrow Z \rightarrow Y,
\end{array}
\right.\\[3mm]
y^n \ci \{z_y^n, x_y^n\} & \Longrightarrow &
\left\{
\begin{array}{lll}
z_y^n \ci x_{yz}^n & \Longrightarrow & Y \rightarrow Z \rightarrow X,\\
x_y^n \ci z_{yx}^n
& \Longrightarrow &
Y \rightarrow X \rightarrow Z
\end{array}
\right.\\[3mm]
z^n \ci \{x_z^n, y_z^n\} & \Longrightarrow &
\left\{
\begin{array}{lll}
x_z^n \ci y_{zx}^n & \Longrightarrow & Z \rightarrow X \rightarrow Y,\\
y_z^n \ci x_{zy}^n & \Longrightarrow & Z \rightarrow Y \rightarrow X.
\end{array}
\right.
\end{array}
\right.
\]
In the multivariate case, PW-LiNG-based DirectLiNGAM determines the
variable order greedily using pairwise exogeneity scores.
Let $\mathcal{U}$ denote the set of variables that have not yet
been ordered. 
The algorithm chooses $i\in\mathcal{U}$ that maximizes
\[
-
\sum_{\substack{j\in\mathcal{U}\\ j\neq i}}
\left\{
\min\bigl(0,\widehat{D}_{ij}\bigr)
\right\}^{2},
\]
where $\widehat{D}_{ij}$ denotes the PW-LiNG score for the ordered pair
$(X_i,X_j)$. The same procedure is repeated on the resulting residuals.

\begin{figure}
\begin{center}
{\setlength{\unitlength}{0.50mm}
\small
\begin{picture}(220,80)(0,-30)
\put(10,40){(a)}
\put(110,40){(b)}
\put(10,5){\circle{10}}
\put(10,5){\makebox(0,0){$X$}}
\put(40,5){\circle{10}}
\put(40,5){\makebox(0,0){$Y$}}
\put(70,5){\circle{10}}
\put(70,5){\makebox(0,0){$Z$}}
\put(100,5){\circle{10}}
\put(100,5){\makebox(0,0){$W$}}
\put(15,5){\vector(1,0){20}}
\put(45,5){\vector(1,0){20}}
\put(75,5){\vector(1,0){20}}

\put(35,20){\dashbox{0.5}(14,14){$e_{XZ}$}}
\put(80,-25){\dashbox{0.5}(14,14){$e_{ZW}$}}
\put(35,-25){\dashbox{0.5}(14,14){$e_Y$}}
\put(40,-10){\vector(0,1){10}}

\put(120,5){\circle{10}}
\put(120,5){\makebox(0,0){$X$}}
\put(150,5){\circle{10}}
\put(150,5){\makebox(0,0){$Y$}}
\put(180,5){\circle{10}}
\put(180,5){\makebox(0,0){$Z$}}
\put(210,5){\circle{10}}
\put(210,5){\makebox(0,0){$W$}}
\put(125,5){\vector(1,0){20}}
\put(155,5){\vector(1,0){20}}
\put(185,5){\vector(1,0){20}}

\put(145,20){\dashbox{0.5}(14,14){$e_{XZ}$}}
\put(175,-25){\dashbox{0.5}(14,14){$e_{YW}$}}

\thicklines
\put(10,25){\color{red}\line(1,0){25}}
\put(50,25){\color{red}\line(1,0){20}}
\put(10,25){\color{red}\vector(0,-1){15}}  
\put(70,25){\color{red}\vector(0,-1){15}}  
\put(70,-15){\color{red}\line(1,0){10}}
\put(95,-15){\color{red}\line(1,0){5}}
\put(70,-15){\color{red}\vector(0,1){15}}  
\put(100,-15){\color{red}\vector(0,1){15}} 

\put(120,25){\color{red}\line(1,0){25}}
\put(160,25){\color{red}\line(1,0){20}}
\put(120,25){\color{red}\vector(0,-1){15}} 
\put(180,25){\color{red}\vector(0,-1){15}} 
\put(150,-15){\color{red}\line(1,0){25}}
\put(190,-15){\color{red}\line(1,0){20}}
\put(150,-15){\color{red}\vector(0,1){15}} 
\put(210,-15){\color{red}\vector(0,1){15}} 
\end{picture}
}
\end{center}
\caption{\label{dw2023}
In Fig.~(a), confounders \(e_{XZ}\) and \(e_{ZW}\) exist
between \(X\) and \(Z\), and between \(Z\) and \(W\), respectively.
Therefore, although identifying the order \(Z\rightarrow W\) may be difficult, the order \(X\rightarrow Y\rightarrow Z\) can still be
correctly identified. In Fig.~(b), the full order
\(X\rightarrow Y\rightarrow Z\rightarrow W\) can be correctly identified.
}
\end{figure}

Since DirectLiNGAM selects upstream variables greedily, it does not
evaluate the global structure of the entire order. Even in the absence
of confounding, a local decision based on finite samples may lead to a
suboptimal order because of estimation error. This limitation becomes
particularly important in the presence of confounding, where exact
independence generally does not hold. Thus, evaluating the entire order
can be advantageous even without confounding, and its advantage becomes
more pronounced when confounding is present.

Wang and Drton \citep{WangDrton2023} proposed a causal discovery
method for bow-free acyclic path diagrams (Fig. \ref{dw2023}). The bow-free condition
excludes the simultaneous presence of a directed edge and a bidirected
edge between the same pair of variables.

\subsection{ICA-LiNGAM}

Independent Component Analysis (ICA; \cite{comon94,HyvarinenOja2000})
aims to represent a random vector
\(X=(X_1,\ldots,X_p)^\top\) as a linear mixture of mutually independent
non-Gaussian components.  That is, ICA assumes that there exist a
nonsingular matrix \(A\in{\mathbb R}^{p\times p}\) and mutually
independent non-Gaussian random variables
\(Z_1,\ldots,Z_p\) such that
$$\left[
\begin{array}{c}
X_1\\
\vdots\\
X_p
\end{array}
\right]
=
A
\left[
\begin{array}{c}
Z_1\\
\vdots\\
Z_p
\end{array}
\right].$$

We define the quantity \citep{watanabe1960information} that measures dependence among 
\(Z_1,\ldots,Z_p\) by 
\begin{equation}\label{eq3}
K
:=
{\mathbb E}
\left[
\log
\frac{
p_Z(Z_1,\ldots,Z_p)
}{
p_1(Z_1)\cdots p_p(Z_p)
}
\right]
=
\sum_{j=1}^p h(Z_j)-h(Z_1,\ldots,Z_p),
\end{equation}
where \(p_Z\) and \(p_j\) denote the joint density of
\((Z_1,\ldots,Z_p)\) and the marginal density of \(Z_j\), respectively.
Here \(h(Z_1,\ldots,Z_p)\) and \(h(Z_j)\) denote the corresponding
differential entropies.  We have
\[
K\geq 0,
\]
with equality if and only if
\(Z_1,\ldots,Z_p\) are mutually independent.  Hence, in principle,
ICA may be formulated as the problem of finding a linear transformation
that minimizes \(K\).

However, the direct computation and minimization of \eqref{eq3} are generally difficult.  Practical ICA algorithms therefore replace the direct evaluation of total correlation by computationally simpler measures of non-Gaussianity.  In particular, they use the negentropy $J(U)$ introduced in \eqref{eq:direct-negentropy}.

After whitening the observed variables, ICA searches for a linear transformation whose components are as mutually independent as possible.  Algorithms such as FastICA perform this optimization by maximizing approximations of the marginal negentropies of the transformed components \citep{HyvarinenOja2000}.

Thus, PW-LiNG and FastICA are based on the same general concept of negentropy, although they use it differently.  
FastICA optimizes a linear transformation so that the resulting components have large non-Gaussianity and are approximately mutually independent.

ICA-LiNGAM applies this ICA idea to a linear structural equation model.
It assumes that
\[
\left[
\begin{array}{c}
X_1\\
\vdots\\
X_p
\end{array}
\right]
=
B
\left[
\begin{array}{c}
X_1\\
\vdots\\
X_p
\end{array}
\right]
+
\left[
\begin{array}{c}
Z_1\\
\vdots\\
Z_p
\end{array}
\right],
\]
where \(Z=(Z_1,\ldots,Z_p)^\top\) consists of mutually independent
non-Gaussian error variables.  If the variables are ordered according
to a causal order, then \(B=(b_{ij})\) is a strictly lower triangular
matrix.  Thus, LiNGAM has an ICA representation with
\[
A=(I-B)^{-1}.
\]
However, this is not a general ICA model with an arbitrary nonsingular
mixing matrix.  Rather, it is an ICA representation restricted by the
triangular structure induced by an acyclic linear structural equation
model.

ICA-LiNGAM therefore relies on ICA algorithms to recover the independent
non-Gaussian error variables and then uses the triangular structure to
identify the causal order.  In this sense, ICA-LiNGAM assumes that the
structural errors are mutually independent, or equivalently, that there
is no latent confounding among them.  The present study keeps the
triangular structural equation model but evaluates the remaining dependence
among the transformed error variables by the information-theoretic quantity
\(K\) in \eqref{eq3}.

\subsection{Gaussian Copula and Its Marginal Likelihood}
\label{subsec:gaussian-copula-marginal}

Let $Q\geq 1$, and let
$X=(X_1,\ldots,X_Q)^\top$ be a random vector.
Let $x_i=(x_{i1},\ldots,x_{iQ})^\top\in{\mathbb R}^Q$,
$i=1,\ldots,n$, be independent realizations of $X$.
We fix the degrees of freedom $\nu>0$ and assume that, for each
$j=1,\ldots,Q$, the marginal distribution of $X_j$ is a univariate
Student $t$ distribution with location $\mu_j$ and scale $\sigma_j>0$:
\[
f_{\nu}(x\mid\mu_j,\sigma_j)
=
\frac{1}{\sigma_j}
\frac{\Gamma\left((\nu+1)/2\right)}
{\Gamma\left(\nu/2\right)\sqrt{\nu\pi}}
\left\{
1+
\frac{1}{\nu}
\left(
\frac{x-\mu_j}{\sigma_j}
\right)^2
\right\}^{-(\nu+1)/2},
\qquad
x\in{\mathbb R}.
\]
Let $T_{\nu}$ denote the distribution function of the standard
$t_{\nu}$ distribution, and define
\[
U_j
=
T_{\nu}\left(
\frac{X_j-\mu_j}{\sigma_j}
\right),
\qquad
Y_j
=
\Phi^{-1}(U_j),
\qquad
j=1,\ldots,Q,
\]
where $\Phi$ is the standard Gaussian distribution function.
For the $i$th observation, the corresponding realizations are
\[
u_{ij}
=
T_{\nu}\left(
\frac{x_{ij}-\mu_j}{\sigma_j}
\right),
\qquad
y_{ij}
=
\Phi^{-1}(u_{ij}).
\]
We assume that
$Y=(Y_1,\ldots,Y_Q)^\top\sim N_Q(0,R)$,
where $R$ is a $Q\times Q$ positive definite correlation matrix,
and each $(y_{i1},\ldots,y_{iQ})$ is the realization of $Y$.
For $u=(u_1,\ldots,u_Q)^\top$ and
$y=(\Phi^{-1}(u_1),\ldots,\Phi^{-1}(u_Q))^\top$, the Gaussian copula
density is
\[
c_R(u)
=
|R|^{-1/2}
\exp\left\{
-\frac{1}{2}
y^\top\left(R^{-1}-I_Q\right)y
\right\}.
\]
The corresponding joint density with $t_\nu$ marginal distributions is
\[
c_R(u_1,\ldots,u_Q)
\prod_{j=1}^{Q}
f_\nu(x_j\mid\mu_j,\sigma_j).
\]
In particular, $R=I_Q$ represents independence among the $Q$ coordinates.

For numerical optimization, put $\eta_j=\log\sigma_j$, and let
${\gamma}$ denote any smooth local coordinates for the free
off-diagonal entries of $R$.  
We write
\[
\theta
=
\left(
\mu_1,\ldots,\mu_Q,
\eta_1,\ldots,\eta_Q,
\gamma^\top
\right)^\top
\in\Theta_Q\subset\mathbb R^{d_Q},
\qquad
d_Q
=
2Q+\frac{Q(Q-1)}{2}
=
\frac{Q(Q+3)}{2}.
\]
The log likelihood based on ${x}_1,\ldots,{x}_n$
is
\begin{equation}
\ell^{n}({\theta})
:={}
-
\frac{n}{2}\log|R|
-
\frac{1}{2}
\sum_{i=1}^{n}
{y}_i^{\top}
\left(R^{-1}-I_Q\right)
{y}_i+
\sum_{i=1}^{n}
\sum_{j=1}^{Q}
\log f_{\nu}(x_{ij}\mid\mu_j,\sigma_j).
\label{eq21}    
\end{equation}
Let $\pi_Q({\theta})$ be a prior density that is positive and
smooth in a neighborhood of the maximum likelihood estimator
\begin{equation}\label{eq22}
\widehat{\theta}^{n}
:=
\mathop{\mathrm{arg\,max}}_{\theta\in\Theta_Q}
\ell^{n}(\theta).
\end{equation}
We define the marginal likelihood by
\[
g^{n}
:=
\int_\Theta
\exp\left\{
\ell^{n}(\theta)
\right\}
\pi_Q(\theta)
\,d\theta
\]
and the observed information matrix per observation by
\[
\widehat{\mathcal{I}}^{n}
:=
-
\frac{1}{n}
\left.
\frac{\partial^2\ell^{n}(\theta)}
{\partial\theta\partial\theta^\top}
\right|_{\theta=\widehat{\theta}^{n}}.
\]
Under the usual regularity conditions, the Laplace expansion gives
$$\log g^{n}
={}
\ell^{n}\left(
\widehat{{\theta}}^{n}
\right)
-
\frac{d_Q}{2}\log n
+
\frac{d_Q}{2}\log(2\pi)
+
\log\pi_Q\left(
\widehat{{\theta}}^{n}
\right)
-
\frac{1}{2}
\log\left|
\widehat{\mathcal{I}}^{n}
\right|
+
O\left(n^{-1}\right).
\label{eq:laplace-log-marginal}$$

For the special case $Q=1$,  the Gaussian copula density is
identically equal to one.  
Put $\eta=\log\sigma$.
Then
$$
\ell^{n}(\mu,\eta)
=
n\left\{
\log\Gamma\left(\frac{\nu+1}{2}\right)
-
\log\Gamma\left(\frac{\nu}{2}\right)
-
\frac{1}{2}\log(\nu\pi)
-
\eta
\right\}
-
\frac{\nu+1}{2}
\sum_{i=1}^{n}
\log\left\{
1+
\frac{e^{-2\eta}(x_i-\mu)^2}{\nu}
\right\}.
$$
Let
\[
(\widehat{\mu},\widehat{\eta})
=
\mathop{\mathrm{arg\,max}}_{(\mu,\eta)}
\ell^{n}(\mu,\eta),
\qquad
\widehat{\sigma}=e^{\widehat{\eta}},
\qquad
\widehat{r}_i
=
\frac{x_i-\widehat{\mu}}{\widehat{\sigma}}.
\]
The entries of the two-dimensional observed information matrix
$\widehat{\mathcal{I}}^{n}$ are obtained directly from the data as
$$\widehat{\mathcal{I}}^n_{\mu\mu}
={}
\frac{\nu+1}{n\widehat{\sigma}^{2}}
\sum_{i=1}^{n}
\frac{\nu-\widehat{r}_i^{2}}
{\left(\nu+\widehat{r}_i^{2}\right)^{2}},
\label{eq:univariate-info-mumu}\quad
\widehat{\mathcal{I}}_{\mu\eta}^n
={}
\frac{2\nu(\nu+1)}{n\widehat{\sigma}}
\sum_{i=1}^{n}
\frac{\widehat{r}_i}
{\left(\nu+\widehat{r}_i^{2}\right)^{2}},
\label{eq:univariate-info-mueta}\quad
\widehat{\mathcal{I}}_{\eta\eta}^n
={}
\frac{2\nu(\nu+1)}{n}
\sum_{i=1}^{n}
\frac{\widehat{r}_i^{2}}
{\left(\nu+\widehat{r}_i^{2}\right)^{2}}.
\label{eq:univariate-info-etaeta}
$$
Since $d_1=2$, the normalized negative log marginal likelihood is
$$
\begin{aligned}
-\frac{1}{n}\log g^{n}
={}&
-\frac{1}{n}\ell^{n}(\widehat{\mu},\widehat{\eta})
+
\frac{\log n}{n}
-
\frac{\log(2\pi)}{n}
-
\frac{1}{n}\log\pi_1(\widehat{\mu},\widehat{\eta})
+
\frac{1}{2n}
\log\left|
\widehat{\mathcal{I}}^{n}
\right|
+
O\left(n^{-2}\right).
\end{aligned}
$$
In particular, once the prior $\pi_1$ has been specified, all terms of
orders $n$, $\log n$, and $1$ in $\log g^{n}$ are computable from the
observed sample. Hence no unspecified $O(1)$ term remains. The omitted
remainder is $O(n^{-1})$ in $\log g^{n}$, or equivalently $O(n^{-2})$
in $-n^{-1}\log g^{n}$.

\section{Bayesian LiNGAM}

DirectLiNGAM \citep{shimizu11} determines the causal order greedily by repeatedly
selecting an upstream variable according to local pairwise scores.
This limitation is often less serious when confounding is absent or
weak. However, DirectLiNGAM does not optimize a criterion over the
entire causal order, and once an incorrect variable is selected, the
decision cannot be corrected at a later stage. This problem becomes
particularly serious as confounding increases, because exact
independence no longer holds at each step and the local scores may lead
to an incorrect causal order.

ICA-LiNGAM \citep{shimizu06}, on the other hand, relies on FastICA \citep{HyvarinenOja2000}, which uses
approximations of marginal negentropy as computational surrogates for
independence. The fundamental ICA criterion is the multivariate mutual
information $K$ in \eqref{eq3}. However, ICA-LiNGAM does not directly
minimize $K$ over causal orders. In particular, under confounding and
with finite samples, optimization of the negentropy-based surrogate
need not yield the order that minimizes the remaining dependence among
the residual variables.

To overcome these two limitations, 
we propose a method, called \emph{Bayesian LiNGAM}, that estimates the
multivariate mutual information $K$ using Bayesian marginal likelihoods
and globally optimizes it over complete causal orders.
Since $K\geq0$, with equality if and only if
the residual variables are mutually independent, it provides an
absolute measure of their remaining dependence. Bayesian LiNGAM
therefore avoids the greedy local decisions of DirectLiNGAM while
directly targeting the original ICA criterion instead of replacing it
by a negentropy-based surrogate. The following subsections define the
proposed criterion, establish its consistency, and describe its global
shortest-path search.

\subsection{Consistent Estimation of Confounding}

For each causal order, we relabel the observed random variables as
$X_1,\ldots,X_p$ and write
\[
X_i
=
\sum_{j=1}^{i-1}\beta_{i,j}X_j+Z_i,
\qquad
 i=1,\ldots,p.
\]
In the absence of confounding, the residual variables
$Z_1,\ldots,Z_p$ can be mutually independent. The causal order is
estimated from observed realizations of $X_1,\ldots,X_p$.

In this paper, the magnitude of confounding is quantified by the
multivariate mutual information $K$ in \eqref{eq3}. As explained in
Section~2.2, ICA seeks components that minimize this quantity. In the
present setting, among the $p!$ possible orders, we select the one that
yields the smallest value of $K$. We use the decomposition
$$K
=
\sum_{j=1}^{p-1}
\mathbb{E}\!\left[
\log
\frac{p(Z_j,\ldots,Z_p)}
     {p(Z_j)p(Z_{j+1},\ldots,Z_p)}
\right]
=
\sum_{j=1}^{p-1}
I(Z_j;Z_{j+1},\ldots,Z_p),
$$
where
\[
I(Z_j;Z_{j+1},\ldots,Z_p)
:=
\mathbb{E}\!\left[
\log
\frac{p(Z_j,\ldots,Z_p)}
     {p(Z_j)p(Z_{j+1},\ldots,Z_p)}
\right].
\]

For a fixed order, let
$x_i=(x_i^{(1)},\ldots,x_i^{(p)})^\top$ and 
$z_i=(z_i^{(1)},\ldots,z_i^{(p)})^\top$,
$i=1,\ldots,n$,
be the observed data and 
the corresponding residual vectors. 
Hence the mutual information can
be estimated from the residuals, and the estimated value of $K$ depends
on the order of $X_1,\ldots,X_p$.
Throughout this subsection, the causal order is fixed and its
dependence is suppressed from the notation. When the order must be
made explicit, we write $Z_j^\tau$, $g_{j:p}^{n,\tau}$,
$g_j^{n,\tau}$, $\widehat I_j(\tau)$, and
$\widehat K_n(\tau)$.

We adopt the Gaussian copula model introduced in
Section~\ref{subsec:gaussian-copula-marginal} because it separates the
Student $t_\nu$ marginal distributions from the dependence structure,
which is represented by the correlation matrix $R$. This construction
allows non-Gaussian marginal distributions while retaining a simple
parametric model of dependence. Moreover, mutual independence among
the $Q$ coordinates is represented by $R=I_Q$, or equivalently, by the
constant copula density $c_R(u)=1$.

For the fixed order, let $g_{j:p}^n$ denote the marginal likelihood
obtained by applying the Gaussian-copula $t_\nu$ model in
Section~\ref{subsec:gaussian-copula-marginal} to the residual variables
remaining immediately before the $j$th variable is selected, and let
$g_j^n$ denote the univariate marginal likelihood of the residual
$Z_j$ selected at that step. Thus, $g_{1:p}^n$ is the marginal
likelihood of the original data, whereas $g_{(j+1):p}^n$ is the
marginal likelihood of the residual variables remaining after the
linear effect of $Z_j$ has been removed. For $j=1,\ldots,p-1$, define
\begin{equation}\label{eq4}
\widehat I_j
:=
\frac{1}{n}
\log
\frac{g_{j:p}^n}{g_j^n g_{(j+1):p}^n}.
\end{equation}

The residual block represented by $g_{j:p}^n$ is transformed into
$(Z_j,\ldots,Z_p)^\top$ by the successive residualizations associated
with the fixed order. This transformation is triangular with unit
diagonal and therefore has determinant one. Hence, the differential
entropy of the residual block represented by $g_{j:p}^n$ is
\[
h(Z_j,\ldots,Z_p),
\]
and the population counterpart of $\widehat I_j$ is
\[
I(Z_j;Z_{j+1},\ldots,Z_p).
\]

Since
\[
g_{p:p}^n=g_p^n,
\]
the corresponding estimator of $K$ is
\[
\widehat K_n
:=
\sum_{j=1}^{p-1}\widehat I_j
=
\frac{1}{n}
\left\{
\log g_{1:p}^n
-
\sum_{j=1}^p\log g_j^n
\right\}.
\]
Since $K(Z_1,\ldots,Z_p)=0$ if and only if the residual variables are
mutually independent, $\widehat K_n$ measures their remaining
dependence.

The main result in this subsection is the consistency of these marginal-likelihood
estimators under the Gaussian-copula $t_\nu$ model.
Fix $\nu>0$. Let $A$ be a nonempty finite index set, put
$Q=|A|$, and let
\[
Z_A=(Z_j:j\in A)^\top
\]
denote a $Q$-dimensional residual vector. For
$z_A=(z_j:j\in A)^\top$, consider
\[
p_A(z_A\mid\theta_A)
=
c_{R_A}(u_A)
\prod_{j\in A}f_\nu(z_j\mid\mu_j,\sigma_j),
\qquad
u_j
=
T_\nu\!\left(\frac{z_j-\mu_j}{\sigma_j}\right),
\]
where $u_A=(u_j:j\in A)^\top$, $c_{R_A}$ is the $Q$-dimensional
Gaussian copula density, $R_A$ is a positive definite correlation
matrix, and $c_{R_A}\equiv1$ when $Q=1$. Let
\[
\theta_A
=
\left(
\mu_A^\top,
\eta_A^\top,
\gamma_A^\top
\right)^\top
\in\Theta_A\subset\mathbb{R}^{d_A},
\qquad
\eta_j=\log\sigma_j,
\qquad
d_A=\frac{Q(Q+3)}{2},
\]
where
$\mu_A=(\mu_j:j\in A)^\top$,
$\eta_A=(\eta_j:j\in A)^\top$, and $\gamma_A$ denotes coordinates
for the free entries of $R_A$.
Let $z_{1,A},\ldots,z_{n,A}$ be independent observations from $p_{0,A}$,
and define
\[
g_A^n
=
\int_{\Theta_A}
\prod_{i=1}^n
p_A(z_{i,A}\mid\theta_A)
\,\pi_A(d\theta_A).
\]

\begin{teiri}[Consistency of the Gaussian-copula $t_\nu$ marginal likelihood]
\label{kei1}

Let $p_{0,A}$ be the true density of $Z_A$. Assume that
\begin{enumerate}
\item there exists $\theta_A^0\in\operatorname{int}(\Theta_A)$ such that
$p_{0,A}(z_A)=p_A(z_A\mid\theta_A^0)$;
\item $\Theta_A$ is compact,
$\theta_A\mapsto\log p_A(z_A\mid\theta_A)$ is continuous for
$p_{0,A}$-almost every $z_A$, and
\[
\mathbb{E}_0\!\left[
\sup_{\theta_A\in\Theta_A}
\left|\log p_A(Z_A\mid\theta_A)\right|
\right]
<\infty;
\]
\item $\pi_A$ is a proper prior that assigns positive mass to every
neighborhood of $\theta_A^0$.
\end{enumerate}
Then
\[
-\frac{1}{n}\log g_A^n
\ \xrightarrow{p}\
h(Z_A).
\]
If the regularity conditions for the Laplace expansion in
Section~\ref{subsec:gaussian-copula-marginal} also hold, the same limit
holds when $g_A^n$ is replaced by its Laplace approximation.
\end{teiri}

For the proof, see Appendix \ref{proof-1}.

\begin{kei}
\label{cor:gaussian-copula-mi-consistency}
Suppose that the assumptions of Theorem~\ref{kei1} hold for all
residual vectors whose marginal likelihoods appear in \eqref{eq4}.
Then, for each $j=1,\ldots,p-1$,
\[
\widehat I_j
\ \xrightarrow{p}\
I(Z_j;Z_{j+1},\ldots,Z_p),
\]
and
\[
\widehat K_n
=
\sum_{j=1}^{p-1}\widehat I_j
\ \xrightarrow{p}\
K(Z_1,\ldots,Z_p).
\]
\end{kei}
\noindent\textit{Proof of Corollary~\ref{cor:gaussian-copula-mi-consistency}.}
By Theorem~\ref{kei1} and the unit-determinant relation described
above,
\[
\frac{1}{n}\log g_{j:p}^n
\ \xrightarrow{p}\
-h(Z_j,\ldots,Z_p),
\]
\[
\frac{1}{n}\log g_j^n
\ \xrightarrow{p}\
-h(Z_j),
\]
and
\[
\frac{1}{n}\log g_{(j+1):p}^n
\ \xrightarrow{p}\
-h(Z_{j+1},\ldots,Z_p).
\]
Hence, for $j=1,\ldots,p-1$,
\[
\begin{aligned}
\widehat I_j
&\ \xrightarrow{p}\
-h(Z_j,\ldots,Z_p)
+h(Z_j)
+h(Z_{j+1},\ldots,Z_p)\\
&=
I(Z_j;Z_{j+1},\ldots,Z_p).
\end{aligned}
\]
Similarly,
\[
\widehat K_n
\ \xrightarrow{p}\
\sum_{j=1}^p h(Z_j)-h(Z_1,\ldots,Z_p)
=
K(Z_1,\ldots,Z_p).
\]
\hfill$\square$

We call the resulting method \emph{Bayesian LiNGAM}, since it uses
Bayesian marginal likelihoods to estimate the ICA criterion $K$ and
select the causal order.

\paragraph{Identifiability.}
For each causal order $\tau$, let $K_\tau$ denote the multivariate
mutual information among the residual variables obtained under
$\tau$. The causal order is identifiable by the proposed criterion if
$K_\tau$ is uniquely minimized by one order. If two or more orders
attain the same minimum, they cannot be distinguished by this
criterion.

For example, if two variables are jointly Gaussian, the regression
residual is independent of the regressor in both directions, and hence
$K_\tau=0$ for both orders. More generally, identifiability requires
that one causal order yield a strictly smaller multivariate mutual
information than every other order. In this paper, we assume identifiability:
\begin{katei}\label{katei1}
There exists a unique causal order $\tau^*$ such that
\[
K_{\tau^*}<K_\tau
\qquad
\text{for every }\tau\neq\tau^*.
\]
\end{katei}

The non-Gaussianity condition is used only to establish the
computational result in Theorem~\ref{teiri2}. Specifically, it
guarantees the linear-ICA separability needed to show that a
competing variable associated with a zero-length edge must coincide,
up to scale, with one of the independent disturbances. This condition
is not imposed on the general model allowing confounding considered
elsewhere in the paper.

\subsection{Search Strategy}

\begin{figure}
\begin{wrapfigure}{r}[10pt]{0.5\textwidth}
{\setlength\unitlength{0.4mm}
\begin{center}
\begin{picture}(80,60)(20,30)
\put(40,70){\framebox(40,10){$x^n,y^n,z^n$}}
\put(60,70){\thicklines\color{red}\line(-4,-1){40}}
\put(60,70){\thicklines\color{blue}\line(0,-1){10}}
\put(60,70){\line(4,-1){40}}
\put(5,50){\framebox(30,10){$y^n_x,z^n_x$}}
\put(45,50){\framebox(30,10){$z_y^n,x_y^n$}}
\put(85,50){\framebox(30,10){$x_z^n,y_z^n$}}
\put(60,50){\line(-2,-1){40}}
\put(60,50){\thicklines\color{blue}\line(2,-1){40}}
\put(20,50){\line(2,-1){40}}
\put(100,50){\line(-2,-1){40}}
\put(20,50){\thicklines\color{red}\line(0,-1){20}}
\put(100,50){\line(0,-1){20}}
\put(10,20){\framebox(20,10){$z_{xy}^n$}}
\put(50,20){\framebox(20,10){$y_{zx}^n$}}
\put(90,20){\framebox(20,10){$x_{yz}^n$}}
\put(20,20){\line(4,-1){40}}
\put(100,20){\line(-4,-1){40}}
\put(60,20){\line(0,-1){10}}
\put(50,0){\framebox(20,10){}}
\end{picture}
\end{center}}
\end{wrapfigure}
\begin{eqnarray*}
K_n(x^n,y^n_x,z^n_{xy})&=&{\color{red}I_n(x^n;y_x^n,z_x^n)+I_n(y_x^n;z_{xy}^n)}\\
K_n(x^n,z_x^n,y_{xz}^n)&=&I_n(x^n;y^n_x,z_x^n)+I_n(z_{x}^n;y_{xz}^n)\\
K_n(y^n,z^n_y,x^n_{yz})&=&{\color{blue}I_n(y^n;z^n_y,x_y^n)+I_n(z_y^n;x_{yz}^n)}\\
K_n(y^n,x_y^n,z_{yx}^n)
&=&
I_n(y^n;z_y^n,x_y^n)
+
I_n(x_y^n;z_{yx}^n)\\
K_n(z^n,x^n_z,y^n_{zx})&=&I_n(z^n;x^n_z,y_z^n)+I_n(x_z^n;y_{zx}^n)\\
K_n(z^n,y_z^n,x_{zy}^n)
&=&
I_n(z^n;x_z^n,y_z^n)
+
I_n(y_z^n;x_{zy}^n)
\end{eqnarray*}
\vspace{1em}
\caption{
\label{fig2}
There are six paths corresponding to the six possible orders.
We compare the sum of the distances from the top node
$\{x^n,y^n,z^n\}$ to the bottom node $\emptyset$
along each path (order).
}
\end{figure}

\begin{figure}
\setlength\unitlength{0.4mm}
\small
\begin{center}
\begin{tabular}{cccc}
\begin{picture}(120,90)(0,15)
\put(10,95){(a)}
\put(40,90){\framebox(40,10){$x^n,y^n,z^n$}}
\put(60,90){\line(-4,-1){40}}
\put(60,90){\line(0,-1){10}}
\put(60,90){\line(4,-1){40}}
\put(5,70){\color{blue}\framebox(30,10){$y^n_x,z^n_x$}}
\put(45,70){\color{blue}\framebox(30,10){$z^n_y,x^n_y$}}
\put(85,70){\color{blue}\framebox(30,10){$x^n_z,y^n_z$}}
\put(60,70){\line(-2,-1){40}}
\put(60,70){\line(2,-1){40}}
\put(20,70){\line(2,-1){40}}
\put(100,70){\line(-2,-1){40}}
\put(20,70){\line(0,-1){20}}
\put(100,70){\line(0,-1){20}}
\put(10,40){\framebox(20,10){$z^n_{xy}$}}
\put(50,40){\framebox(20,10){$y^n_{zx}$}}
\put(90,40){\framebox(20,10){$x^n_{yz}$}}
\put(20,40){\line(4,-1){40}}
\put(100,40){\line(-4,-1){40}}
\put(60,40){\line(0,-1){10}}
\put(50,20){\framebox(20,10){}}
\end{picture}&
\begin{picture}(120,90)(0,15)
\put(10,95){(b)}
\put(40,90){\framebox(40,10){$x^n,y^n,z^n$}}
\put(60,90){\line(-4,-1){40}}
\put(60,90){\line(0,-1){10}}
\put(60,90){\line(4,-1){40}}
\put(5,70){\framebox(30,10){$y^n_x,z^n_x$}}
\put(45,70){\color{blue}\framebox(30,10){$z^n_y,x^n_y$}}
\put(85,70){\color{blue}\framebox(30,10){$x^n_z,y^n_z$}}
\put(60,70){\line(-2,-1){40}}
\put(60,70){\line(2,-1){40}}
\put(20,70){\line(2,-1){40}}
\put(100,70){\line(-2,-1){40}}
\put(20,70){\line(0,-1){20}}
\put(100,70){\line(0,-1){20}}
\put(10,40){\color{blue}\framebox(20,10){$z^n_{xy}$}}
\put(50,40){\color{blue}\framebox(20,10){$y^n_{zx}$}}
\put(90,40){\framebox(20,10){$x^n_{yz}$}}
\put(20,40){\line(4,-1){40}}
\put(100,40){\line(-4,-1){40}}
\put(60,40){\line(0,-1){10}}
\put(50,20){\framebox(20,10){}}\end{picture}&
\begin{picture}(120,90)(0,15)
\put(10,95){(c)}
\put(40,90){\framebox(40,10){$x^n,y^n,z^n$}}
\put(60,90){\line(0,-1){10}}
\put(60,90){\line(4,-1){40}}
\put(5,70){\framebox(30,10){$y^n_x,z^n_x$}}
\put(45,70){\color{blue}\framebox(30,10){$z^n_y,x^n_y$}}
\put(85,70){\color{blue}\framebox(30,10){$x^n_z,y^n_z$}}
\put(60,70){\line(-2,-1){40}}
\put(60,70){\line(2,-1){40}}
\put(20,70){\line(2,-1){40}}
\put(100,70){\line(-2,-1){40}}
\put(100,70){\line(0,-1){20}}
\put(10,40){\framebox(20,10){$z^n_{xy}$}}
\put(50,40){\color{blue}\framebox(20,10){$y^n_{zx}$}}
\put(90,40){\framebox(20,10){$x^n_{yz}$}}
\put(100,40){\line(-4,-1){40}}
\put(60,40){\line(0,-1){10}}
\put(50,20){\color{red}\framebox(20,10){}}
\put(20,40){\color{red}\line(4,-1){40}}
\put(20,70){\color{red}\line(0,-1){20}}
\put(60,90){\color{red}\line(-4,-1){40}}
\end{picture}\\
\begin{picture}(120,90)(0,15)
\put(10,95){(d)}
\put(40,90){\framebox(40,10){$x^n,y^n,z^n$}}
\put(60,90){\line(0,-1){10}}
\put(60,90){\line(4,-1){40}}
\put(5,70){\framebox(30,10){$y^n_x,z^n_x$}}
\put(45,70){\color{blue}\framebox(30,10){$z^n_y,x^n_y$}}
\put(85,70){\color{blue}\framebox(30,10){$x^n_z,y^n_z$}}
\put(60,70){\line(-2,-1){40}}
\put(60,70){\line(2,-1){40}}
\put(20,70){\color{red}\line(2,-1){40}}
\put(100,70){\line(-2,-1){40}}
\put(100,70){\line(0,-1){20}}
\put(10,40){\color{blue}\framebox(20,10){$z^n_{xy}$}}
\put(50,40){\framebox(20,10){$y^n_{zx}$}}
\put(90,40){\framebox(20,10){$x^n_{yz}$}}
\put(100,40){\line(-4,-1){40}}
\put(20,40){\line(4,-1){40}}
\put(60,40){\color{red}\line(0,-1){10}}
\put(50,20){\color{red}\framebox(20,10){}}
\put(20,70){\line(0,-1){20}}
\put(60,90){\color{red}\line(-4,-1){40}}
\end{picture}&
\begin{picture}(120,90)(0,15)
\put(10,95){(e)}
\put(40,90){\framebox(40,10){$x^n,y^n,z^n$}}
\put(60,90){\line(0,-1){10}}
\put(60,90){\line(4,-1){40}}
\put(5,70){\framebox(30,10){$y^n_x,z^n_x$}}
\put(45,70){\framebox(30,10){$z^n_y,x^n_y$}}
\put(85,70){\color{blue}\framebox(30,10){$x^n_z,y^n_z$}}
\put(60,70){\line(-2,-1){40}}
\put(60,70){\line(2,-1){40}}
\put(20,70){\line(2,-1){40}}
\put(100,70){\line(-2,-1){40}}
\put(100,70){\line(0,-1){20}}
\put(10,40){\color{blue}\framebox(20,10){$z^n_{xy}$}}
\put(50,40){\color{blue}\framebox(20,10){$y^n_{zx}$}}
\put(90,40){\color{blue}\framebox(20,10){$x^n_{yz}$}}
\put(100,40){\line(-4,-1){40}}
\put(20,40){\line(4,-1){40}}
\put(60,40){\line(0,-1){10}}
\put(50,20){\framebox(20,10){}}\put(20,70){\line(0,-1){20}}
\put(60,90){\line(-4,-1){40}}
\end{picture}&
\begin{picture}(120,90)(0,15)
\put(10,95){(f)}
\put(40,90){\framebox(40,10){$x^n,y^n,z^n$}}
\put(60,90){\line(0,-1){10}}
\put(60,90){\line(4,-1){40}}
\put(5,70){\framebox(30,10){$y^n_x,z^n_x$}}
\put(45,70){\color{blue}\framebox(30,10){$z^n_y,x^n_y$}}
\put(85,70){\framebox(30,10){$x^n_z,y^n_z$}}
\put(60,70){\line(-2,-1){40}}
\put(60,70){\line(2,-1){40}}
\put(20,70){\line(2,-1){40}}
\put(100,70){\line(-2,-1){40}}
\put(100,70){\line(0,-1){20}}
\put(10,40){\color{blue}\framebox(20,10){$z^n_{xy}$}}
\put(50,40){\color{blue}\framebox(20,10){$y^n_{zx}$}}
\put(90,40){\color{blue}\framebox(20,10){$x^n_{yz}$}}
\put(100,40){\line(-4,-1){40}}
\put(20,40){\line(4,-1){40}}
\put(60,40){\line(0,-1){10}}
\put(50,20){\framebox(20,10){}}\put(20,70){\line(0,-1){20}}
\put(60,90){\line(-4,-1){40}}
\end{picture}
\end{tabular}
\end{center}
\caption{
The shortest path search for finding the best order among $X,Y,Z$.
\label{fig3}
}
\end{figure}

To obtain the optimal solution of Bayesian LiNGAM,
we construct a diagram based on the residuals
and solve a shortest path problem.
Figure~\ref{fig2} illustrates the determination of the order among $X,Y,Z$.
For a residual sample vector $a^n$ and a collection of residual sample
vectors $b_1^n,\ldots,b_q^n$, let
\[
I_n(a^n;b_1^n,\ldots,b_q^n)
\]
denote the estimator $\widehat I_j$ in \eqref{eq4}, with its sample
arguments displayed explicitly. For a complete path, let $K_n$ denote
the corresponding value of $\widehat K_n$, namely, the sum of its
edge lengths.
First, note that 
the residuals associated with a given node do not depend on the path
used to reach that node.
For example, $z_{xy}^n = z_{yx}^n$ holds.
The red path represents the order
$X \rightarrow Y \rightarrow Z$.
When the sum of the Bayesian mutual information estimators
$I_n(x^n; y_x^n, z_x^n)$ and $I_n(y_x^n; z_{xy}^n)$
is minimized among the six possible paths,
the minimum value
$K_n(x^n, y_x^n, z_{xy}^n)$
is obtained.
To ensure nonnegative edge lengths, negative finite-sample estimates
are replaced by zero.

In Figure~\ref{fig3}, we first compute the lengths of the edges from the top node
$\{x^n,y^n,z^n\}$ to the three nodes
$\{y_x^n,z_x^n\}$, $\{z_y^n,x_y^n\}$, and $\{x_z^n,y_z^n\}$ as
\[
d(y_x^n,z_x^n)=I_n(x^n;y_x^n,z_x^n),\quad
d(z_y^n,x_y^n)=I_n(y^n;z_y^n,x_y^n),\quad
d(x_z^n,y_z^n)=I_n(z^n;x_z^n,y_z^n),
\]
where $d(\cdot)$ denotes the cumulative distance from the top node
$\{x^n,y^n,z^n\}$ to the corresponding node.
We then close the top node $\{x^n,y^n,z^n\}$ and open the three nodes
$\{y_x^n,z_x^n\}$, $\{z_y^n,x_y^n\}$, and $\{x_z^n,y_z^n\}$
(Figure~\ref{fig3}(a)).

Suppose that $d(y_x^n,z_x^n)$ is the smallest among the three.
We next compute $I_n(y_x^n;z_{xy}^n)$ and $I_n(z_x^n;y_{zx}^n)$, which yield
\begin{equation}\label{eq17}
d(z_{xy}^n)
:=
d(y_x^n,z_x^n)+I_n(y_x^n;z_{xy}^n)
\end{equation}
and
\[
d(y_{zx}^n)
:=
d(y_x^n,z_x^n)+I_n(z_x^n;y_{zx}^n),
\]
respectively.
We then close the node $\{y_x^n,z_x^n\}$ and open the nodes
$\{z_{xy}^n\}$ and $\{y_{zx}^n\}$
(Figure~\ref{fig3}(b)).

If $d(z_{xy}^n)$ is the smallest value in Figure~\ref{fig3}(b),
then the shortest path corresponds to the order
$X \rightarrow Y \rightarrow Z$
(Figure~\ref{fig3}(c)).
If instead $d(y_{zx}^n)$ is the smallest,
then the shortest path corresponds to
$X \rightarrow Z \rightarrow Y$
(Figure~\ref{fig3}(d)).

On the other hand, if $d(z_y^n,x_y^n)$ is the smallest in Figure~\ref{fig3}(b),
we compute $I_n(z_y^n;x_{yz}^n)$ and $I_n(x_y^n;z_{xy}^n)$ to obtain
\[
d(x_{yz}^n)
:=
d(z_y^n,x_y^n)+I_n(z_y^n;x_{yz}^n)
\]
and
\begin{equation}\label{eq18}
d(z_{xy}^n)
:=
d(z_y^n,x_y^n)+I_n(x_y^n;z_{xy}^n).
\end{equation}
We then close the node $\{z_y^n,x_y^n\}$ and open the nodes
$\{x_{yz}^n\}$ and $\{z_{xy}^n\}$.
Since the values obtained in \eqref{eq17} and \eqref{eq18} may differ,
we resolve this conflict by retaining the smaller value;
that is, 
we update $d(z_{xy}^n)$ to the smaller of the two values
whenever \eqref{eq18} is smaller
(Figure~\ref{fig3}(e)).

Finally, if $d(x_z^n,y_z^n)$ is the smallest in Figure~\ref{fig3}(b),
we arrive at the configuration shown in Figure~\ref{fig3}(f),
where the values of $d(y_{zx}^n)$ conflict.
In this case, the shorter distance from the top node
$\{x^n,y^n,z^n\}$ to the node $\{y_{zx}^n\}$ is selected.
By continuing this procedure, we eventually obtain the distance $d(\emptyset)$
of the shortest path from the top node $\{x^n,y^n,z^n\}$
to the bottom node $\emptyset$.

The same procedure can be applied to an arbitrary number $p$ of variables,
and is not limited to the case $p=3$.

\begin{teiri}\rm\label{teiri2}
Suppose that no confounding exists, that Assumption~\ref{katei1}
holds, and that the residual variables
$Z_1^{\tau^*},\ldots,Z_p^{\tau^*}$ have finite positive variances,
with at least $p-1$ of them being non-Gaussian.
Let $I_n(\cdot;\cdot)$ be a consistent estimator of each mutual
information used as an edge length. Then, with probability tending to
one as $n\to\infty$, the OPEN/CLOSED search expands only the nodes on
the path corresponding to $\tau^*$. Consequently, Bayesian LiNGAM
requires
\[
\frac{p(p+1)}{2}-1
\]
mutual information evaluations. This equals the number of
candidate-variable evaluations performed by DirectLiNGAM.
\end{teiri}

For the proof, see Appendix~\ref{proof-2}.


\subsection{Global Optimization without High-Dimensional Mutual Information Estimation}

This subsection shows that direct estimation of high-dimensional mutual
information can be avoided by using only univariate residual marginal
likelihoods. 
The dynamic-programming formulation provides an alternative
to the zero-corrected shortest-path formulation when some finite-sample
mutual information estimates are negative.

Let $\tau$ be a causal order. For this order, let
$g_{j:p}^{n,\tau}$ 
and 
$g_j^{n,\tau}$
denote the marginal likelihood of the residual
variables remaining at step $j$ and the
univariate marginal likelihood of the residual corresponding to
$X_{\tau_j}$. Then, we have
\[
\widehat I_j(\tau)
=
\frac{1}{n}
\log
\frac{
g_{j:p}^{n,\tau}
}{
g_j^{n,\tau}g_{(j+1):p}^{n,\tau}
},
\qquad
j=1,\ldots,p-1.
\]
Since
\[
g_{p:p}^{n,\tau}=g_p^{n,\tau},
\]
summing these terms gives
\[
\begin{aligned}
\widehat K_n(\tau)
=
\sum_{j=1}^{p-1}\widehat I_j(\tau)
=
\frac{1}{n}
\left\{
\log g_{1:p}^{n}
-
\sum_{j=1}^{p}\log g_j^{n,\tau}
\right\}.
\end{aligned}
\]
Thus, all intermediate multivariate marginal likelihoods
$g_{2:p}^{n,\tau},\ldots,g_{(p-1):p}^{n,\tau}$ cancel.

Note that 
$g_{1:p}^{n}$ is the marginal likelihood of the original data and is
common to all causal orders.
 Therefore,
\[
\begin{aligned}
\widehat{\tau}
=
\mathop{\mathrm{arg\,min}}_{\tau}
\widehat K_n(\tau)=
\mathop{\mathrm{arg\,min}}_{\tau}
\sum_{j=1}^{p}
\left\{
-\frac{1}{n}\log g_j^{n,\tau}
\right\}
\end{aligned}
\]
Consequently, the globally optimal order can be obtained using only the
univariate marginal likelihoods of the successive residuals.

This optimization can also be written as dynamic programming. Let
$V=\{1,\ldots,p\}$, and let $S\subseteq V$ be the set of variables that
have already been selected. For $i\notin S$, let $g_{i\mid S}^{n}$
denote the univariate marginal likelihood of the residual obtained by
regressing $X_i$ on the variables indexed by $S$, and define
\begin{equation}\label{eq51}
w_n(S,i)
=
-\frac{1}{n}\log g_{i\mid S}^{n}.
\end{equation}
Then, the minimizing order is obtained from $D_n(V)$, and
\[
\mathop{\mathrm{min}}_{\tau}\widehat K_n(\tau)
=
\frac{1}{n}\log g_{1:p}^{n}+D_n(V),
\]
where $D_n(S)$ with $S\subseteq V$ is defined by
\begin{equation}\label{eq52}
D_n(S):=
\left\{
\begin{array}{ll}
0,&\displaystyle S=\emptyset\\
\min_{i\in S}
\left\{
D_n(S\setminus\{i\})
+
w_n(S\setminus\{i\},i)
\right\}, & S\not=\emptyset
\end{array}
\right.
\end{equation}

A direct estimate of
$
I(Z_j;Z_{j+1},\ldots,Z_p)
$
requires the estimation of a joint distribution whose dimension
increases with $p-j+1$. When the number of variables is large relative
to the sample size, such high-dimensional estimation is affected by the
curse of dimensionality. The cancellation above avoids this difficulty:
although the population criterion is the multivariate mutual information
$K$, its global minimization requires only univariate marginal
likelihoods.

PW-LiNG also avoids direct estimation of bivariate mutual information by
using univariate negentropy approximations. However, it uses pairwise
directional scores and determines the order greedily. ICA-LiNGAM uses
negentropy approximations to estimate an ICA transformation. In
contrast, Bayesian LiNGAM retains the multivariate mutual information
$K$ itself as the population criterion and optimizes it globally over
complete causal orders. Since
\[
K\geq0,
\]
with equality if and only if the residual variables are mutually
independent, $K$ gives an absolute measure of the remaining dependence.
Thus, Bayesian LiNGAM combines one-dimensional marginal-likelihood
estimation with global optimization of the original ICA criterion.

\section{Implementation}

This section describes the implementation and numerical evaluation of
the method developed in Section~3. Section~4.1 presents the hybrid
Dijkstra--dynamic-programming search used in practice. Section~4.2
reports preliminary experiments examining the finite-sample behavior
underlying Theorem~\ref{teiri2} and comparing the proposed
edge-length estimator with KSG. Finally, Section~4.3 compares Bayesian
LiNGAM with DirectLiNGAM and ICA-LiNGAM.


\subsection{Hybrid Search of Dijkstra and DP}
\label{subsec:implementation}

We first describe the implementation of the order search used in the
numerical experiments. Although mutual information is nonnegative at
the population level, an edge length estimated from finite data can be
negative. Since Dijkstra's algorithm requires nonnegative edge lengths,
the estimated edge lengths cannot always be used directly.

A possible remedy is to replace a negative edge length by zero.
However, this truncation changes the complete-path criterion and may
create many artificial zero-cost edges. In particular, several
candidate paths may then have exactly the same score, which can
deteriorate causal-order recovery. We therefore do not truncate
genuinely negative edge lengths.

\paragraph{Reweighting by maximized likelihoods.}

Let $V=\{1,\ldots,p\}$, and let $S\subseteq V$ denote the set of
variables already selected. 
For each node $S$, let
\[
\phi_n(S)
=
\frac{1}{n}\ell_S(\widehat{\theta}_S),
\]
where $\ell_S(\widehat{\theta}_S)$ is the maximized log likelihood $l^n(\widehat{\theta}^n)$ of
the multivariate residual block at node $S$ under the model introduced
in Section~\ref{subsec:gaussian-copula-marginal}. We use the reweighted
edge length
\[
\widetilde{w}_n(S,i)
=
w_n(S,i)
+
\phi_n(S)
-
\phi_n(S\cup\{i\}),
\]
where $w_n(S,i)$ is defined in \eqref{eq51} for $i\not\in S$.
The purpose of this reweighting is to reduce the occurrence of negative
edge lengths. 
Indeed, since $\phi_n(S)>
\phi_n(S\cup\{i\})$, we see that
$\widetilde{w}_n(S,i)<0$ occurs less often than
${w}_n(S,i)<0$ does.
Note that it does not change the minimizing complete
order. To see this, let
$\tau=(\tau_1,\ldots,\tau_p)$ be a complete order and put
\[
S_j=\{\tau_1,\ldots,\tau_j\},
\qquad
S_0=\emptyset.
\]
Then,
\[
\begin{aligned}
\sum_{j=1}^{p}
\widetilde{w}_n(S_{j-1},\tau_j)
&=
\sum_{j=1}^{p}
w_n(S_{j-1},\tau_j)
+
\sum_{j=1}^{p}
\left\{
\phi_n(S_{j-1})-\phi_n(S_j)
\right\}
\\
&=
\sum_{j=1}^{p}
w_n(S_{j-1},\tau_j)
+
\phi_n(\emptyset)-\phi_n(V).
\end{aligned}
\]
The last two terms are common to all complete orders. Hence, all
intermediate multivariate maximized log likelihoods cancel
telescopically, and
\[
\mathop{\mathrm{arg\,min}}_{\tau}
\sum_{j=1}^{p}
\widetilde{w}_n(S_{j-1},\tau_j)
=
\mathop{\mathrm{arg\,min}}_{\tau}
\sum_{j=1}^{p}
w_n(S_{j-1},\tau_j).
\]
Thus, replacing the intermediate multivariate marginal likelihoods by
maximized likelihoods is only a reweighting of the subset graph. It may
change the order in which nodes are expanded and the resulting
computation time, but it does not change the globally optimal complete
order.

\paragraph{Dijkstra--dynamic-programming hybrid.}

When all the reweighted edge lengths are nonnegative, up to a prescribed
numerical tolerance, we apply Dijkstra's algorithm
for the shortest path problem to the subset graph.
The algorithm can then terminate without enumerating all $2^p$ subset
nodes.

If a genuinely negative edge length remains, we instead solve the same
optimization problem by exact dynamic programming. 
Since the subset graph is acyclic, the recursion \eqref{eq52} remains valid even
when some edge lengths are negative. The minimizing predecessors
recover the globally optimal causal order. Thus, no genuine negative
edge length is modified: Dijkstra's algorithm is used when the edge
lengths are nonnegative, whereas exact dynamic programming is used as a
fallback when a negative value occurs.

\paragraph{Evaluation measures.}
Let
$\tau^0=(\tau_1^0,\ldots,\tau_p^0)$
and 
$
\widehat{\tau}
=
(\widehat{\tau}_1,\ldots,\widehat{\tau}_p)
$
be the true order and an estimated order. For a variable $i$, let
$r_{\widehat{\tau}}(i)$ denote its position in
$\widehat{\tau}$.

The exact-order recovery indicator is defined by
\[
A_{\mathrm{exact}}
\left(
\widehat{\tau},\tau^0
\right)
=
\mathbf{1}
\left\{
\widehat{\tau}=\tau^0
\right\}.
\]
Thus, exact-order recovery equals one only when the entire estimated
order agrees with the true order, and equals zero if even one pair of
variables is incorrectly ordered.

The pairwise-order recovery score is defined by
\[
A_{\mathrm{pair}}
\left(
\widehat{\tau},\tau^0
\right)
=
\frac{1}{\binom{p}{2}}
\sum_{1\leq j<k\leq p}
\mathbf{1}
\left\{
r_{\widehat{\tau}}(\tau_j^0)
<
r_{\widehat{\tau}}(\tau_k^0)
\right\}
=
1-
\frac{
d_{\mathrm{K}}
\left(
\widehat{\tau},\tau^0
\right)
}{
\binom{p}{2}
},
\]
where
$d_{\mathrm{K}}(\widehat{\tau},\tau^0)$ is the Kendall distance,
namely, the number of pairs whose relative order is incorrect.
Therefore, $A_{\mathrm{pair}}=1$ if and only if the complete order is
correct, whereas $A_{\mathrm{pair}}=0$ for the reverse order. Its
expected value is $1/2$ for a uniformly random order.

For repeated experiments, the exact-order and 
pairwise-order recovery rates are the averages of
$A_{\mathrm{exact}}$ and $A_{\mathrm{pair}}$, respectively, over all
replications. Exact-order recovery is a strict all-or-nothing measure,
whereas pairwise-order recovery quantifies how close the estimated
order is to the true order even when exact recovery is not achieved.

\subsection{Preliminary Experiments}

We first examine the finite-sample search behavior underlying
Theorem~\ref{teiri2}. We generated data from a linear non-Gaussian
structural equation model with $p=10$, coefficient magnitude
$\beta=0.4$, and mutually independent $t_{10}$ error variables. No
confounding was introduced. The sample size was varied over
\[
n\in\{300,600,1000\},
\]
and each condition was repeated $100$ times. The variable labels were
randomly permuted in each replication.

For every data set, we applied the OPEN/CLOSED shortest-path search and
also audited the result by exact dynamic programming. We recorded the
exact-order recovery rate, pairwise-order accuracy, whether Dijkstra's
algorithm completed without exact-DP fallback, the number of expanded
nodes, and the number of evaluated edges. The complete subset graph has
$2^{10}=1024$ nodes and $5120$ edges. A search that
follows only one complete path expands $p+1=11$ nodes and
traverses $55$ edges in the implementation. The last transition
has zero cost, so this corresponds to
\[
\frac{p(p+1)}{2}-1=54
\]
nontrivial mutual-information evaluations in Theorem~\ref{teiri2}.

\begin{table}[t]
\centering
\small
\caption{Finite-sample behavior of the shortest-path search without confounding. Search counts are conditional on completion without exact-DP fallback and are reported as mean (standard deviation).}
\label{tab:theorem2-preliminary}
\begin{tabular}{c c c c c c c}
\hline
$n$ & Exact & Pairwise & Repair-free & Expanded & Evaluated & Single-path \\
 & recovery & accuracy & Dijkstra & nodes & edges & rate \\
\hline
300 & 0.490 & 0.863 & 0.950 & 721.7 (134.3) & 3937.7 (601.9) & 0.000 \\
600 & 0.900 & 0.974 & 0.990 & 353.4 (109.9) & 2125.2 (581.9) & 0.000 \\
1000 & 0.960 & 0.994 & 1.000 & 142.3 (57.7) & 917.6 (336.6) & 0.000 \\
\hline
\end{tabular}
\end{table}

\begin{figure}[t]
\centering
\includegraphics[width=0.6\textwidth]{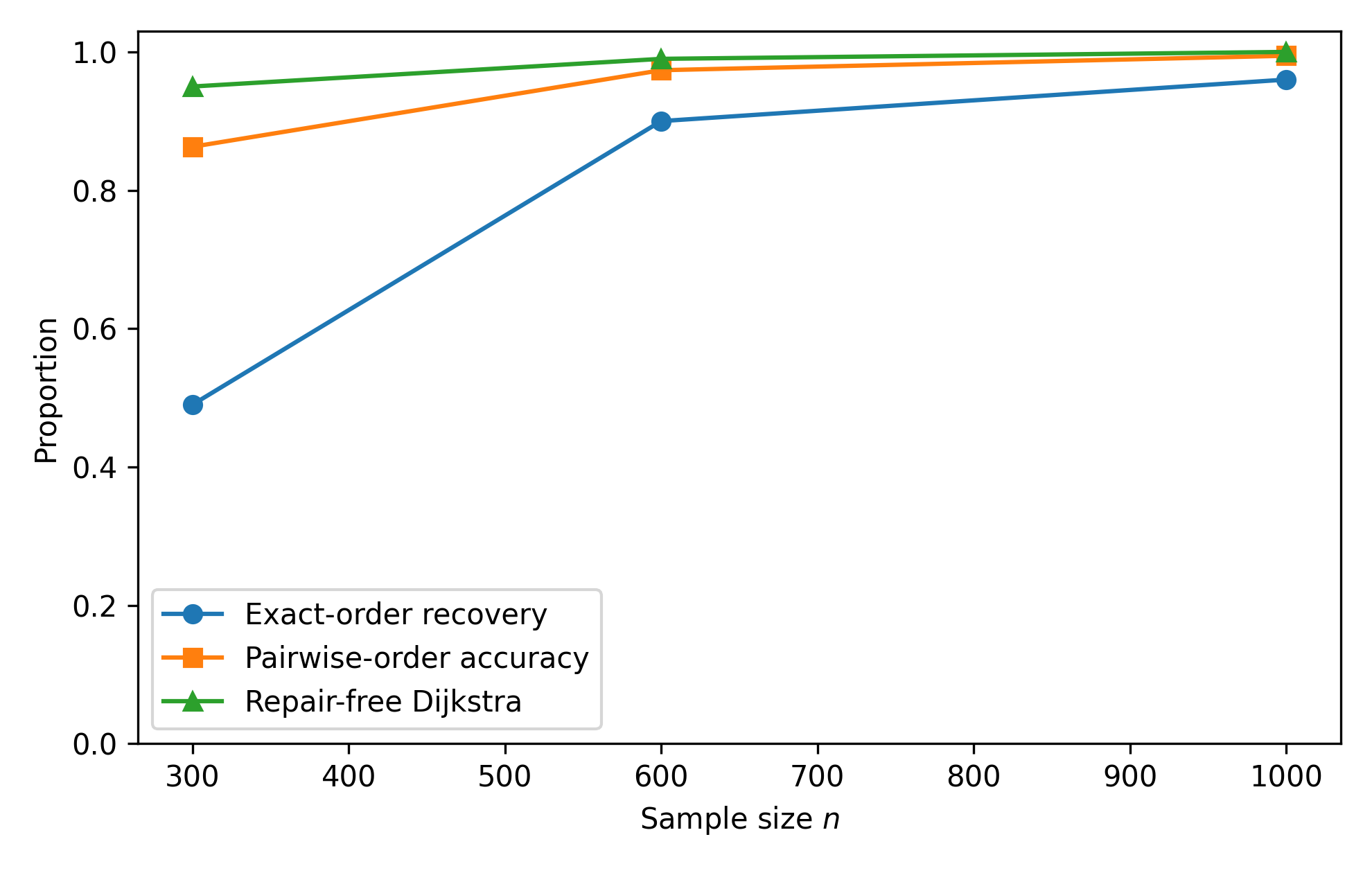}
\caption{Order recovery and repair-free completion in the no-confounding
experiment. Pairwise-order accuracy is one minus the normalized Kendall
distance.}
\label{fig:theorem2-order-recovery}
\end{figure}

\begin{figure}[t]
\centering
\includegraphics[width=0.6\textwidth]{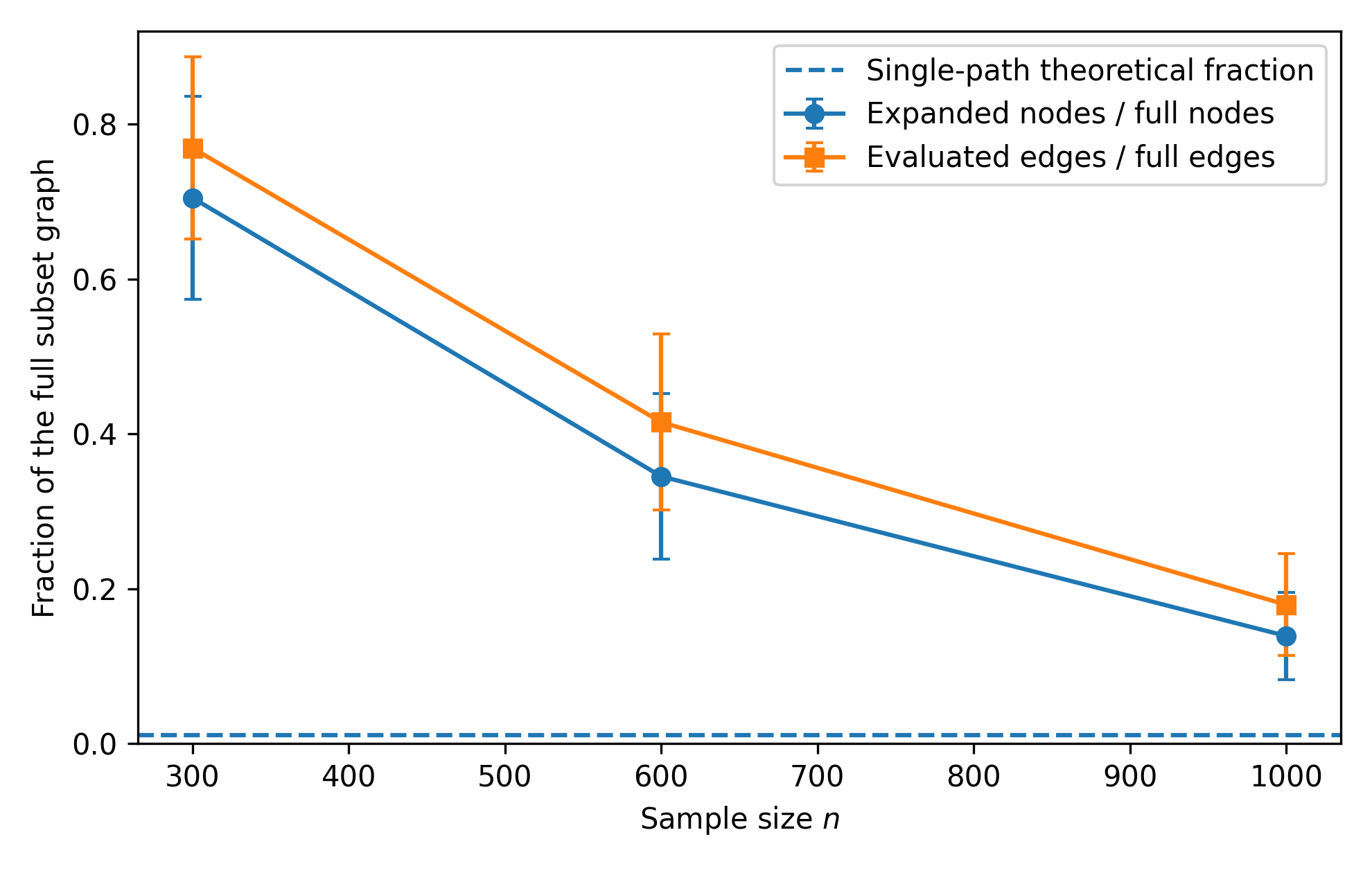}
\caption{Search complexity relative to the full subset graph. Error bars
show one standard deviation across repair-free Dijkstra runs. The dashed
line is the theoretical fraction for a single complete path.}
\label{fig:theorem2-search-complexity}
\end{figure}

Dijkstra's solution agreed with exact dynamic programming in both order
and score in all $300$ data sets. The exact-order recovery rate increased
from $0.49$ at $n=300$ to $0.90$ at $n=600$ and $0.96$ at $n=1000$.
The corresponding pairwise-order accuracies were $0.863$, $0.974$, and
$0.994$. Exact-DP fallback was required in five replications at $n=300$,
one replication at $n=600$, and no replication at $n=1000$.

The search became substantially more concentrated as the sample size
increased. Conditional on repair-free completion, the mean number of
expanded nodes decreased from $721.7$ at $n=300$ to $353.4$ at $n=600$
and $142.3$ at $n=1000$. The mean number of evaluated edges decreased
from $3937.7$ to $2125.2$ and then to $917.6$. For the $100$ matched
replication seeds, both counts decreased monotonically across the three
sample sizes in 98 replications.

These results support the qualitative mechanism behind
Theorem~\ref{teiri2}: in the absence of confounding, increasing the
sample size makes the shortest-path search concentrate around a small
number of low-cost paths. However, none of the present runs attained
the exact single-path counts of 11 nodes and 55 edges.
Thus, the experiment should be interpreted as preliminary finite-sample
evidence for concentration, rather than as empirical attainment of the
asymptotic count in Theorem~\ref{teiri2}.

\paragraph{Comparison with direct KSG estimation.}

We next examine whether the one-versus-rest mutual information terms
used as edge lengths can be estimated directly using the KSG estimator
\citep{Kraskov}. To isolate the effect of mutual information estimation,
we used the same OPEN/CLOSED shortest-path search for both methods and
changed only the edge-length estimator: the Bayesian marginal-likelihood
estimator or the KSG estimator. The KSG estimates were computed using
the Type-I implementation in the Python package \texttt{infomeasure}
\citep{buth2025-infomeasure}, with $k=5$, the maximum norm, and natural
logarithms.

We considered
\[
p\in\{10,20\},
\qquad
n=1000,
\qquad
\beta=0.8,
\]
with $t_{10}$ error variables. Dependence was introduced between a single
pair of error variables, and its strength was varied over
\[
\theta_0\in\{0,0.2,0.4,0.6,0.8\}.
\]
For each value of $\theta_0$, the experiment was repeated $10$ times
when $p=10$ and $5$ times when $p=20$. Negative finite-sample mutual
information estimates were replaced by zero in both methods.

\begin{figure}[t]
\centering
\begin{minipage}{0.48\textwidth}
\centering
\includegraphics[width=\linewidth]{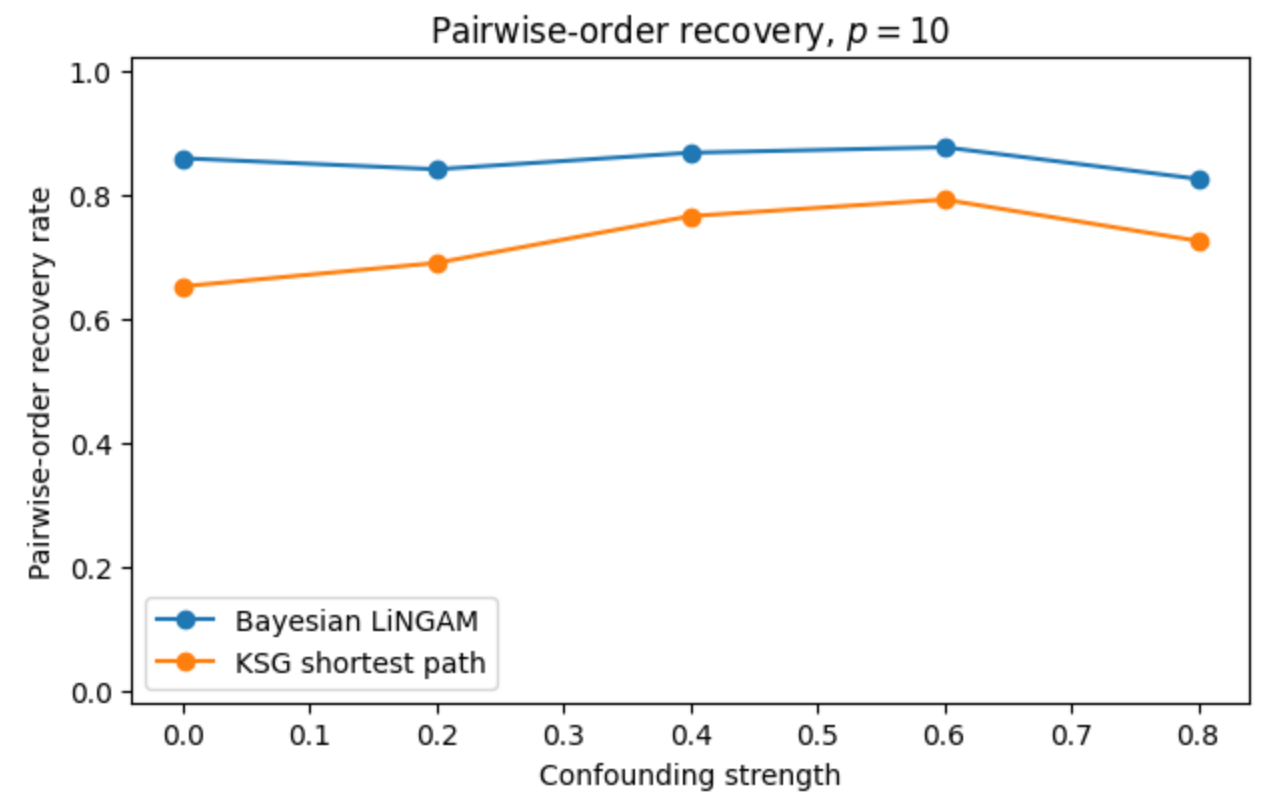}
\end{minipage}
\hfill
\begin{minipage}{0.48\textwidth}
\centering
\includegraphics[width=\linewidth]{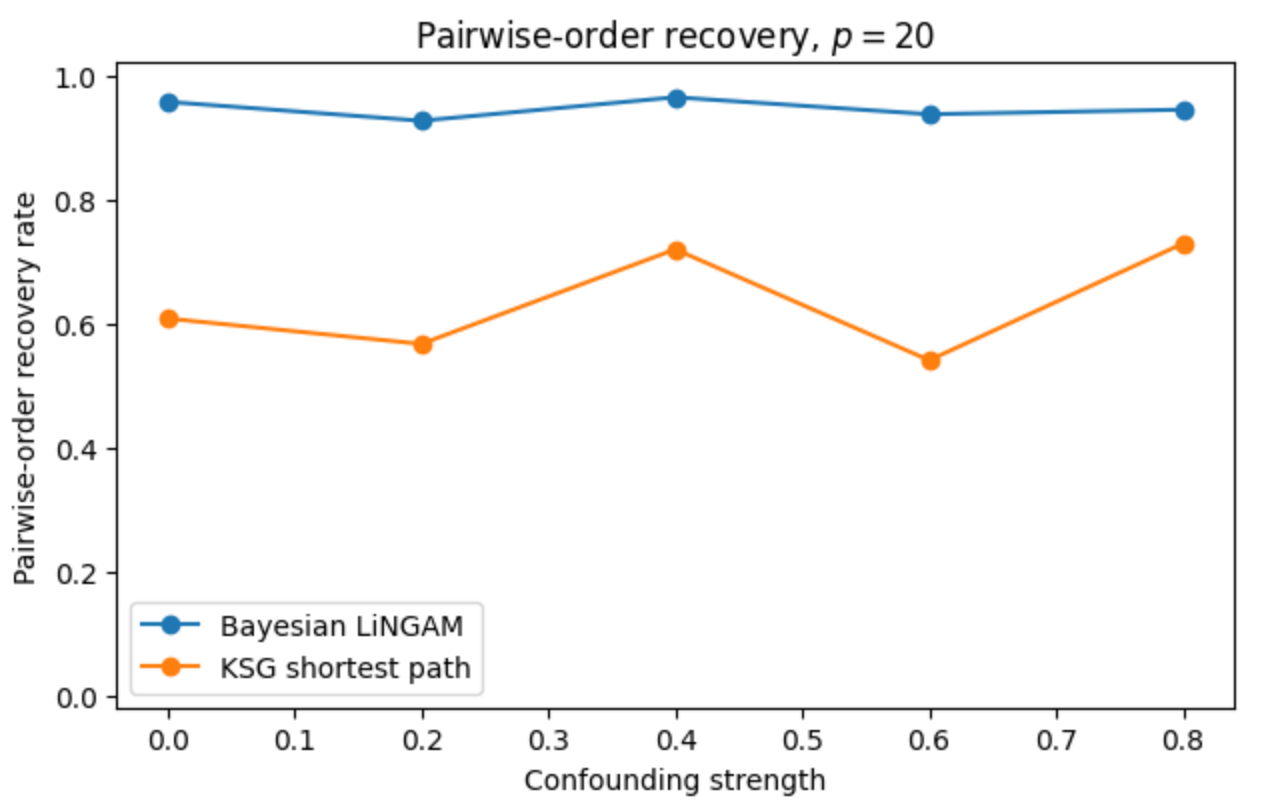}
\end{minipage}
\caption{
Pairwise-order recovery rates of the Bayesian marginal-likelihood
shortest path and the KSG shortest path. Pairwise-order recovery is
defined as one minus the normalized Kendall distance. The left and
right panels correspond to $p=10$ and $p=20$, respectively. Both
methods use the same global shortest-path search and differ only in
the estimation of the edge lengths.
}
\label{fig:bayes-ksg-pairwise}
\end{figure}

For $p=10$, the pairwise-order recovery rate of the Bayesian method
ranged from $0.824$ to $0.876$, whereas that of KSG ranged from
$0.651$ to $0.791$. Thus, the Bayesian method performed better for
all values of $\theta_0$, although the difference was moderate.

The difference became much more pronounced for $p=20$. The
pairwise-order recovery rate of the Bayesian method remained between
$0.926$ and $0.964$, whereas that of KSG ranged only from $0.540$ to
$0.728$. In particular, even without confounding, namely, at
$\theta_0=0$, the recovery rates were $0.957$ for the Bayesian method
and $0.607$ for KSG. Thus, the deterioration of KSG cannot be
attributed only to confounding.

The exact-order recovery rates, which are not shown in
Figure~\ref{fig:bayes-ksg-pairwise}, lead to the same conclusion. For
$p=10$, the exact-order recovery rate of the Bayesian method ranged
from $0$ to $0.4$, whereas that of KSG was at most $0.1$. For $p=20$,
KSG failed to recover the exact order in every condition, whereas the
Bayesian method recovered it in some replications.

The computational difference was also substantial. Averaged over the
five values of $\theta_0$, the mean computation times for $p=10$ were
approximately $0.46$ seconds for the Bayesian method and $4.38$
seconds for KSG. For $p=20$, the corresponding mean computation times
were approximately $2.10$ seconds and $86.44$ seconds, respectively.

The observed difference therefore reflects edge-length estimation
rather than the search strategy. KSG directly estimates mutual
information between one residual variable and a high-dimensional
residual vector, which can be statistically unstable and
computationally expensive as $p$ grows; its deterioration at
$\theta_0=0$ shows that confounding is not the sole cause. These
preliminary results indicate that direct KSG estimation is not well
suited to the present edge lengths. By contrast, the proposed
formulation retains the multivariate mutual information $K$ as the
population criterion while avoiding its direct high-dimensional
estimation through cancellation of the intermediate multivariate
marginal likelihoods, which was discussed in Section 3.3.



\subsection{Comparison with DirectLiNGAM and ICA-LiNGAM}
\label{subsec:comparison-lingam}

We finally compare the proposed complete-order procedure with
DirectLiNGAM and ICA-LiNGAM. In this subsection, the proposed
procedure is implemented using the standardized score described in
Appendix~\ref{app:standardized-score} and is denoted by Bayesian
LiNGAM in the figures and tables. The purpose of the experiment is to
compare the recovery of the complete procedure with the standard
implementations of DirectLiNGAM and ICA-LiNGAM under several
structured-confounding conditions. Because both the ordering criterion
and the search strategy differ between the methods, this comparison
does not isolate the effect of global optimization alone.

\paragraph{Data-generating process.}

Before randomly permuting the observed variables, we generated data from
the linear chain
\[
X_1=Z_1,
\qquad
X_j=\beta X_{j-1}+Z_j,
\qquad
j=2,\ldots,p,
\qquad
\beta=0.4.
\]
The basic disturbance variables
$\varepsilon_1,\ldots,\varepsilon_p$ and all latent common factors
$H_1,H_2,\ldots$ were mutually independent and followed the
$t_{10}$ distribution. In the no-confounding condition, we set
\[
Z_j=\varepsilon_j,\qquad j=1,\ldots,p.
\]
In a confounded condition, each selected pair $(a_r,b_r)$ was assigned
an independent latent common factor $H_r$, and the corresponding
disturbances were generated as
\[
Z_{a_r}=\varepsilon_{a_r}+\gamma H_r,
\qquad
Z_{b_r}=\varepsilon_{b_r}+\gamma H_r.
\]
For variables outside the selected pairs, we retained
$Z_j=\varepsilon_j$. Since the three variables in the preceding display
have the same variance,
\[
\operatorname{Corr}(Z_{a_r},Z_{b_r})
=\frac{\gamma^2}{1+\gamma^2}.
\]
Thus, $\gamma=0.4$ and $\gamma=0.8$ correspond to correlations of
approximately $0.138$ and $0.390$, respectively.

The confounded pairs were defined according to their positions in the
true causal order before the observed columns were randomly permuted. We
used
\[
\mathcal P_{\mathrm{adj}}(p)
=
\left\{
(2r-1,2r):
1\leq r\leq \left\lfloor\frac{p}{2}\right\rfloor
\right\}
\]
for adjacent confounding and
\[
\mathcal P_{\mathrm{nonadj}}(p)
=
\left\{
(2r-1,2r+2):
1\leq r\leq
\left\lfloor\frac{p-2}{2}\right\rfloor
\right\}
\]
for nonadjacent confounding. In the adjacent design, each confounded
pair is also connected by a directed edge in the chain. It therefore
contains both a directed edge and a latent common cause and forms a bow.
In the nonadjacent design, no confounded pair is joined by a direct edge,
and the resulting mixed graph is bow-free. The numbers of confounded
pairs in the adjacent designs were $2$, $5$, and $7$ for
$p=5$, $10$, and $15$, whereas those in the nonadjacent designs were
$1$, $4$, and $6$. Consequently, these designs represent two distinct
structured-confounding patterns rather than a controlled comparison in
which only pair adjacency is changed.

We considered
\[
p\in\{5,10,15\},
\qquad
\frac{n}{p}\in\{10,20,40\},
\qquad
\gamma\in\{0,0.4,0.8\}.
\]
The corresponding sample sizes were $n\in\{50,100,200\}$ for $p=5$,
$n\in\{100,200,400\}$ for $p=10$, and
$n\in\{150,300,600\}$ for $p=15$. Each condition was repeated $100$
times. In every replication, the columns of the generated data matrix
were randomly permuted, and recovery was evaluated against the
generating causal order expressed in the permuted variable labels.

When $\gamma>0$, a confounded disturbance is a sum of independent
$t_{10}$ variables and is generally not itself $t_{10}$. The confounded
conditions therefore assess finite-sample robustness to marginal-model
misspecification as well as to dependent disturbances. The consistency
result in Section~3, which assumes the stated model conditions, is not
invoked for these confounded simulations.

\paragraph{Comparison methods and evaluation measures.}

The implementations of DirectLiNGAM and ICA-LiNGAM in version 1.13.0
of the Python package \texttt{lingam} were used
\citep{IkeuchiEtAl2023}. DirectLiNGAM was run with
\texttt{measure='pwling'}, whereas ICA-LiNGAM was run with
\texttt{max\_iter=1000}. No prior knowledge of the causal order was
supplied to either method.
We used the exact-order and pairwise-order recovery rates defined in
Section~4.1. 

\begin{figure}
\centering
\setlength{\tabcolsep}{2pt}

\begin{tabular}{@{}ccc@{}}
\textbf{No confounding}
&
\textbf{Adjacent, $\gamma=0.8$}
&
\textbf{Nonadjacent, $\gamma=0.8$}
\\[0.2em]

\multicolumn{3}{c}{\textbf{$p=5$}}
\\[-0.4em]

\includegraphics[width=0.31\linewidth]
{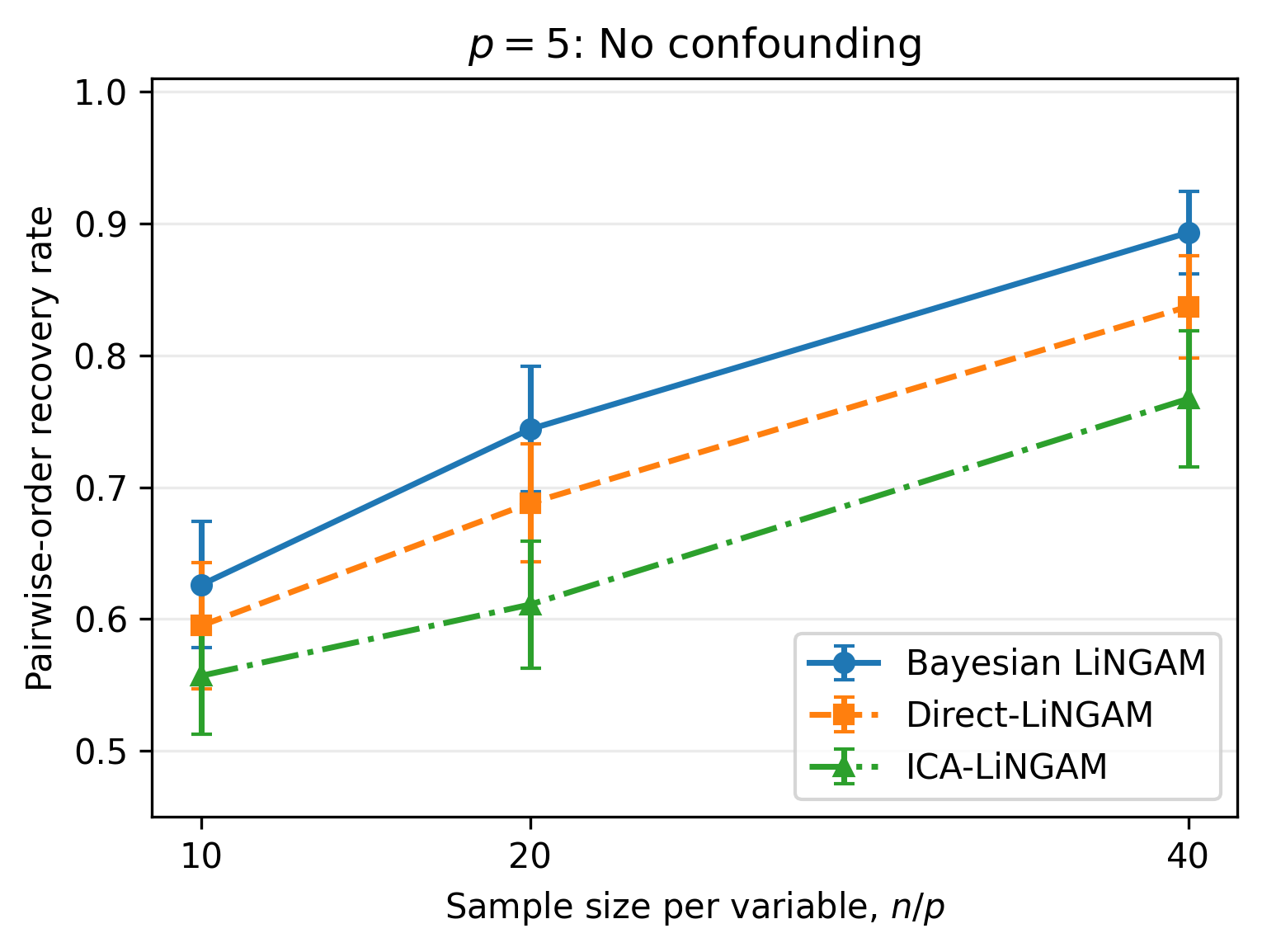}
&
\includegraphics[width=0.31\linewidth]
{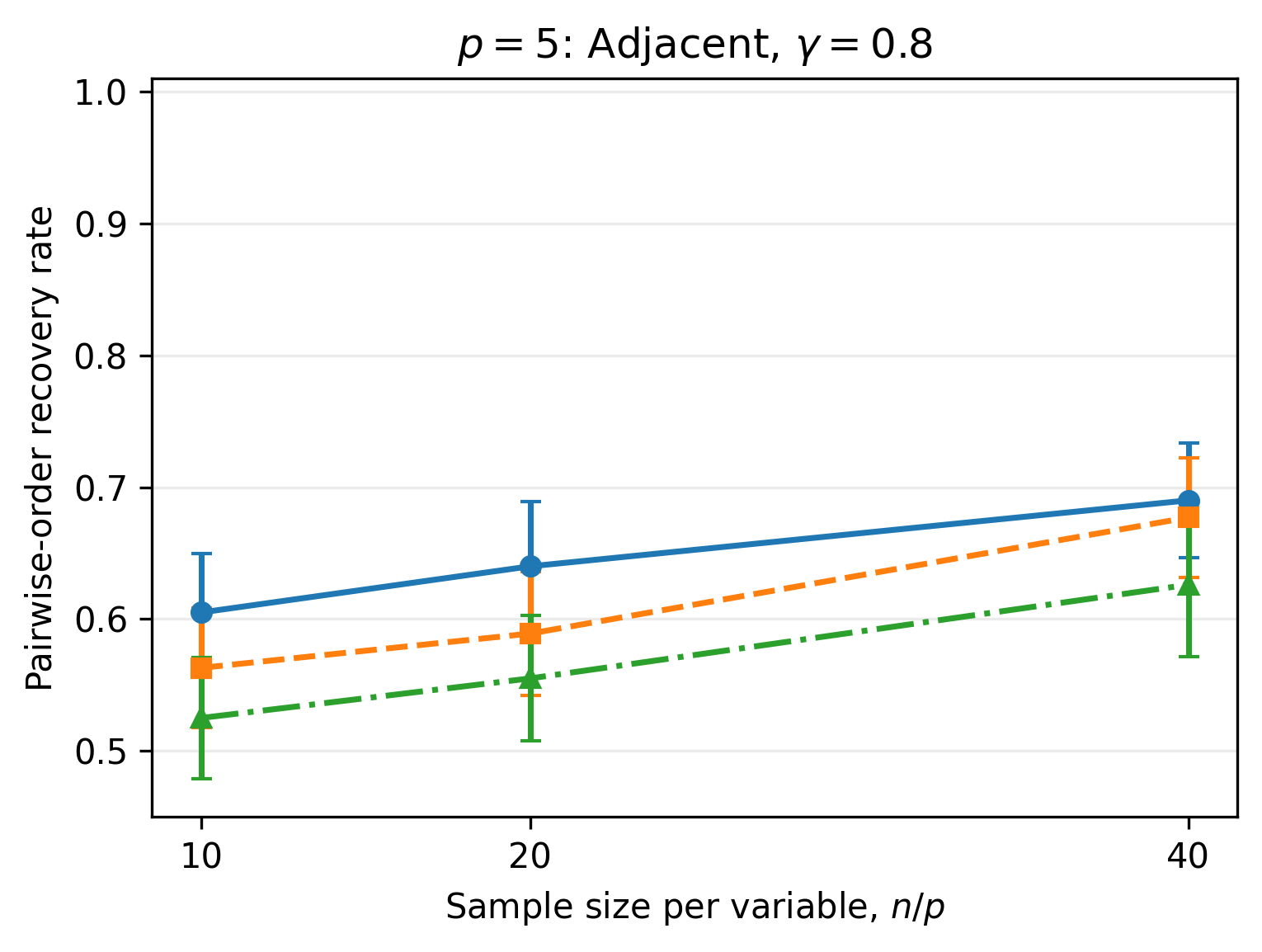}
&
\includegraphics[width=0.31\linewidth]
{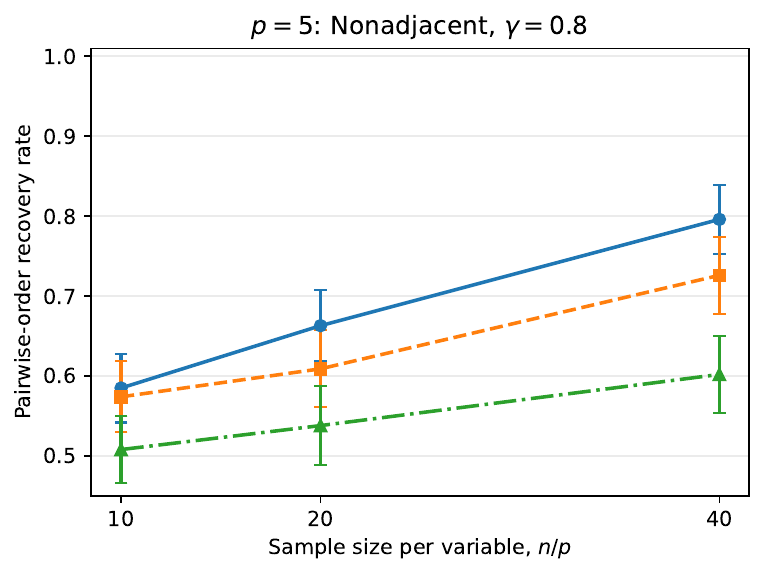}
\\[0.5em]

\multicolumn{3}{c}{\textbf{$p=10$}}
\\[-0.4em]

\includegraphics[width=0.31\linewidth]
{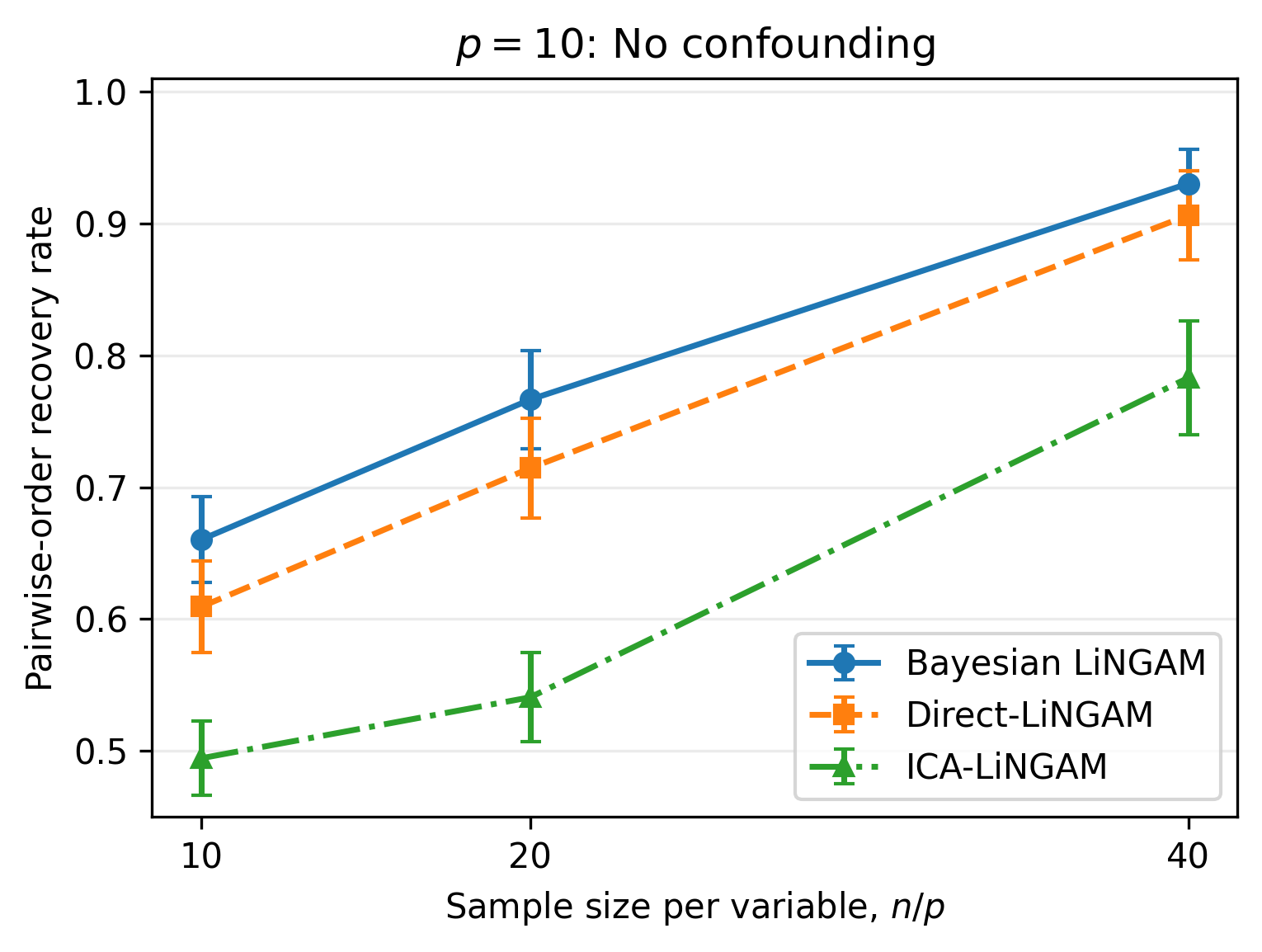}
&
\includegraphics[width=0.31\linewidth]
{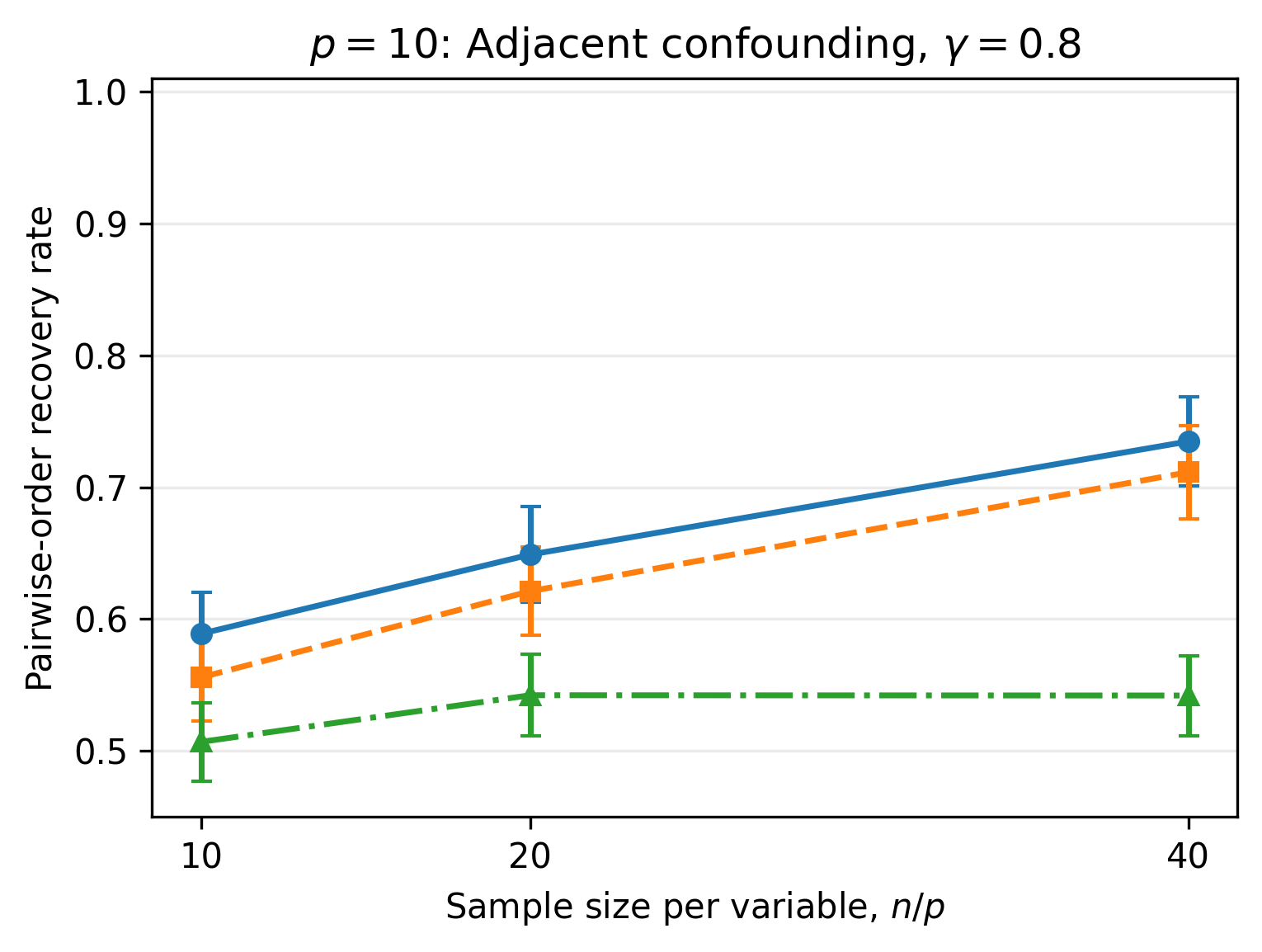}
&
\includegraphics[width=0.31\linewidth]
{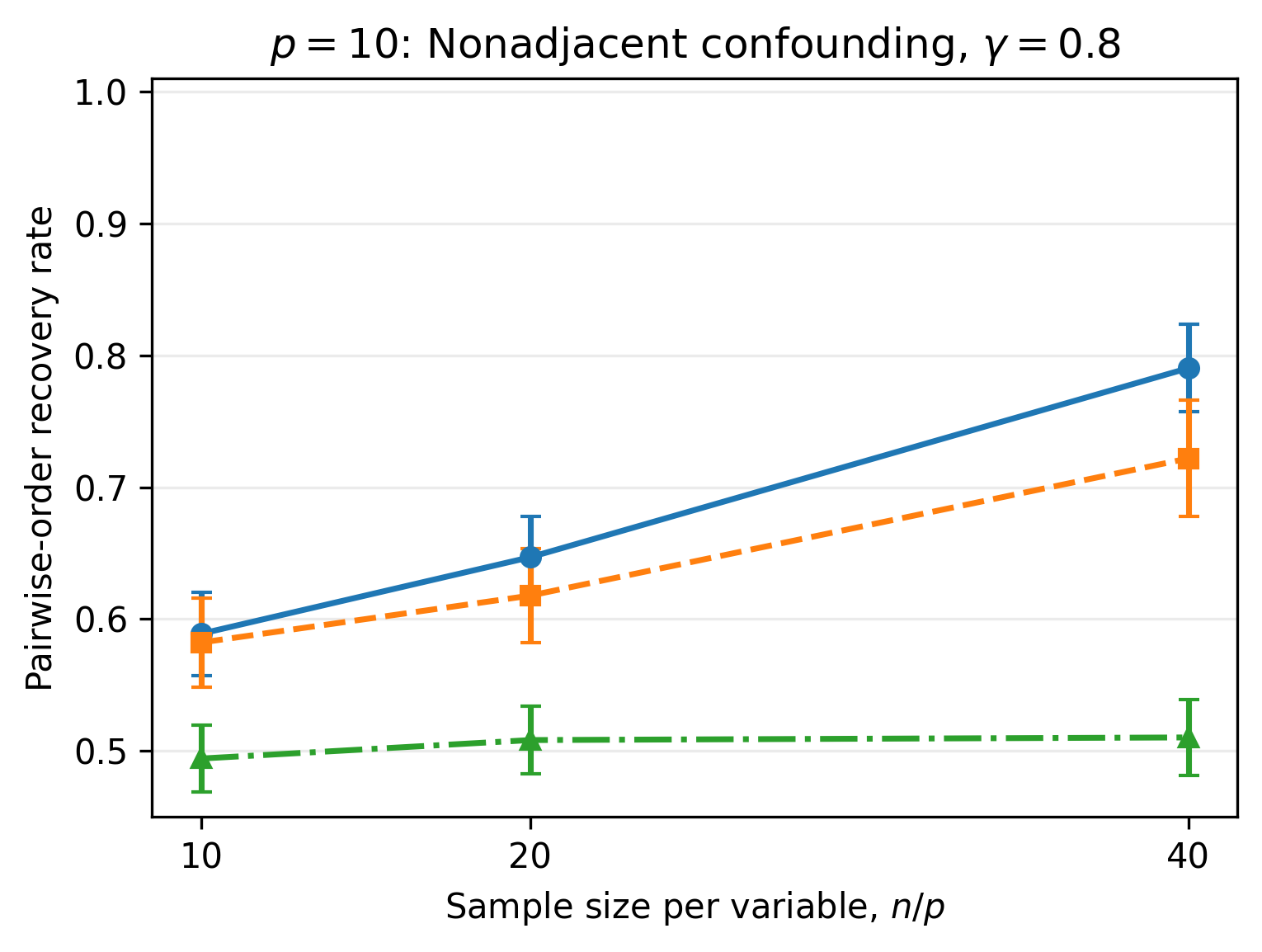}
\\[0.5em]

\multicolumn{3}{c}{\textbf{$p=15$}}
\\[-0.4em]

\includegraphics[width=0.31\linewidth]
{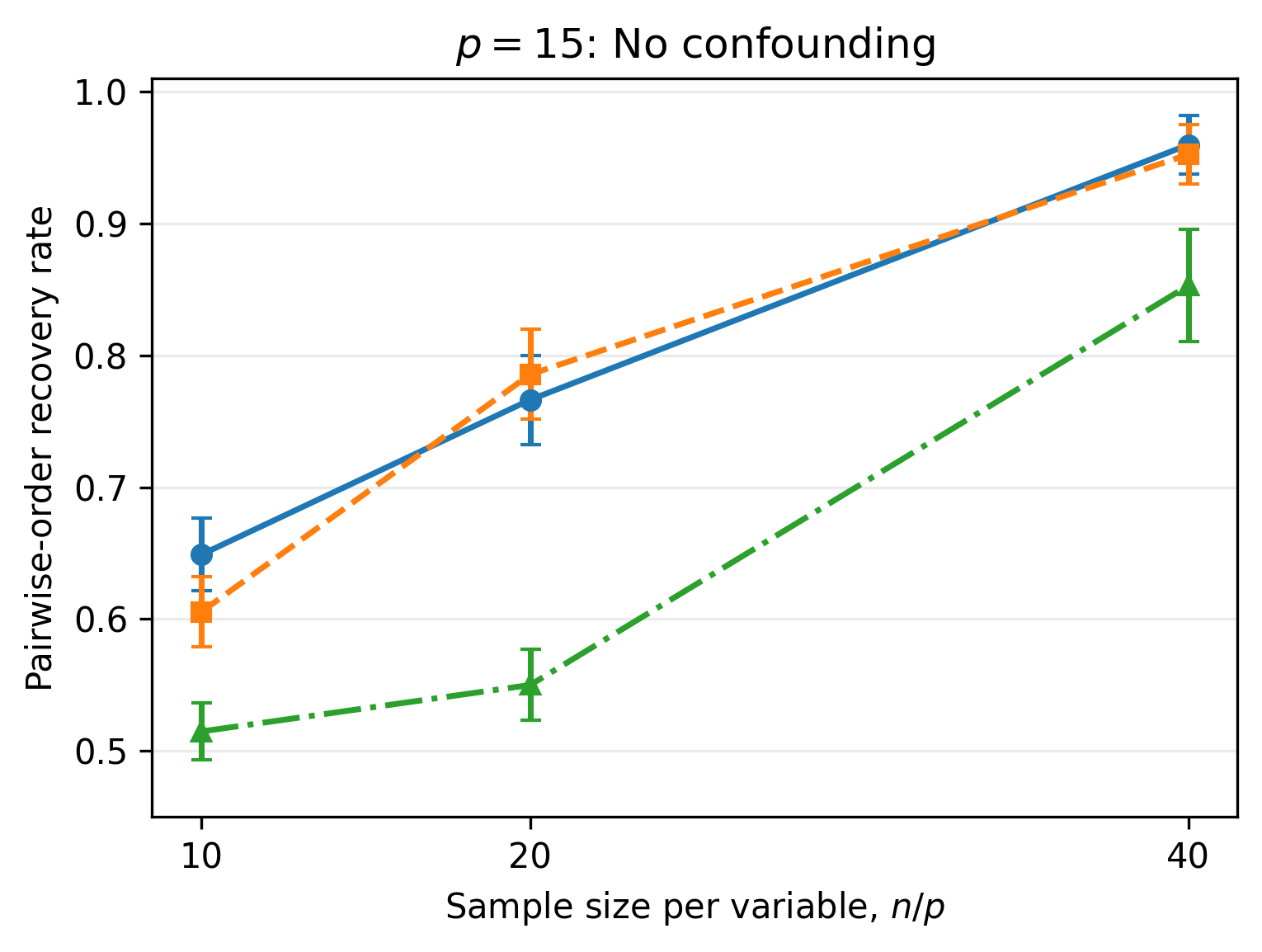}
&
\includegraphics[width=0.31\linewidth]
{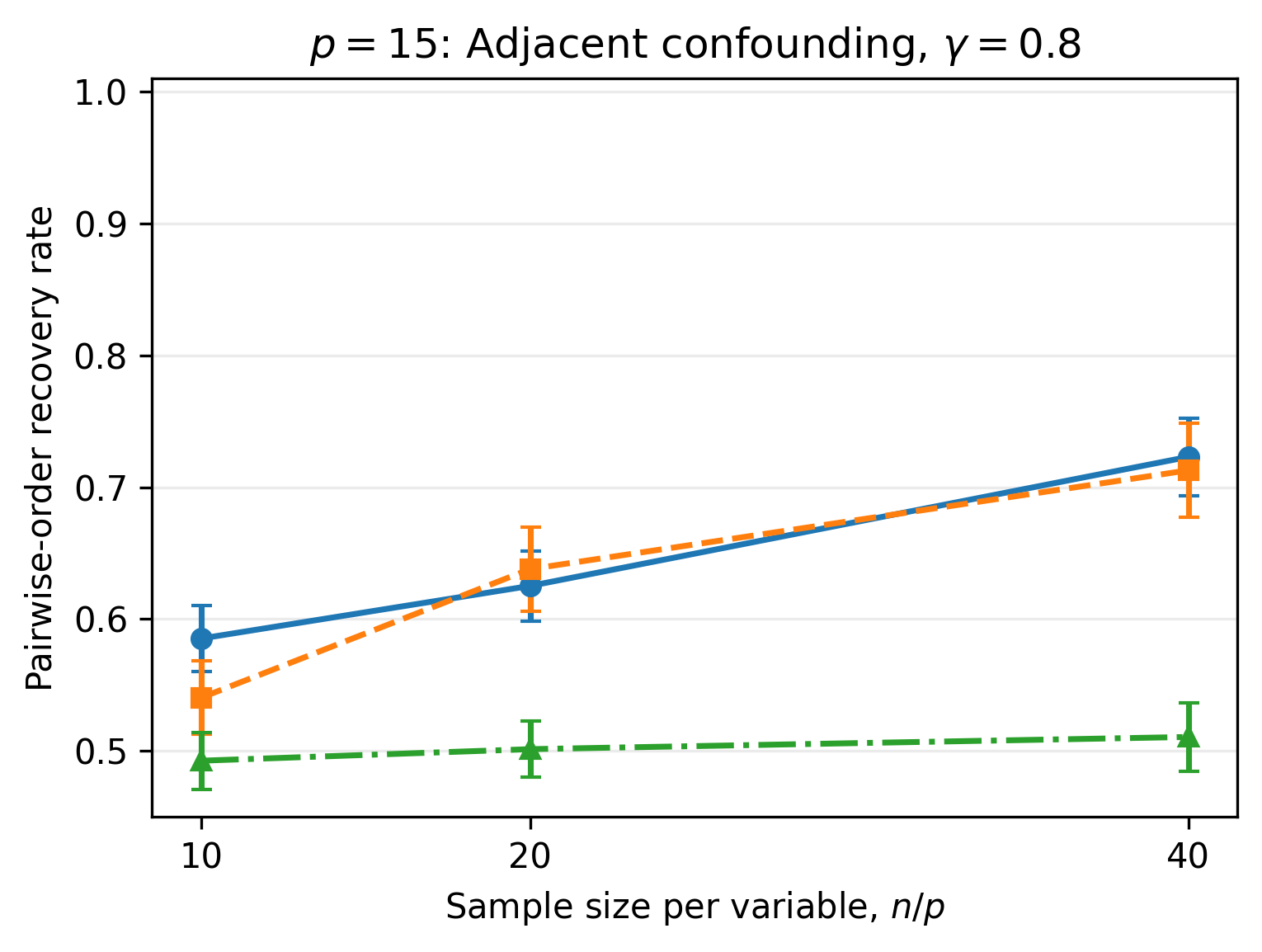}
&
\includegraphics[width=0.31\linewidth]
{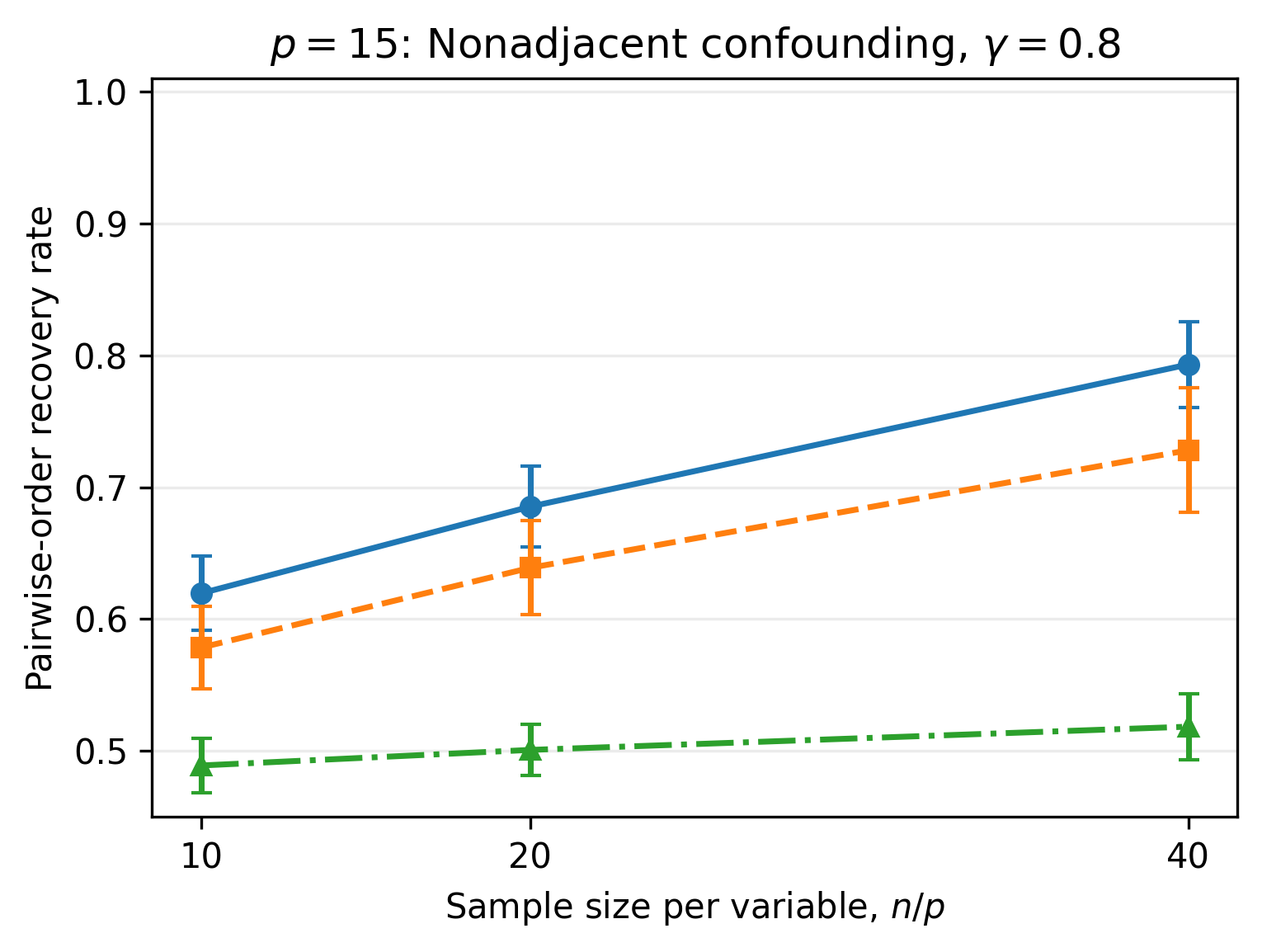}

\end{tabular}

\caption{
Pairwise-order recovery rates of Bayesian LiNGAM, Direct-LiNGAM, and
ICA-LiNGAM. The upper, middle, and lower rows correspond to $p=5$,
$p=10$, and $p=15$, respectively. The columns show no confounding,
adjacent confounding with $\gamma=0.8$, and nonadjacent confounding
with $\gamma=0.8$. The horizontal axis is the sample size per
variable, $n/p$. Error bars show 95\% confidence intervals based on
100 replications.
}
\label{fig:comparison-pairwise}
\end{figure}

\begin{figure}
\centering
\setlength{\tabcolsep}{2pt}

\begin{tabular}{@{}ccc@{}}
\textbf{No confounding}
&
\textbf{Adjacent, $\gamma=0.8$}
&
\textbf{Nonadjacent, $\gamma=0.8$}
\\[0.2em]

\multicolumn{3}{c}{\textbf{$p=5$}}
\\[-0.4em]

\includegraphics[width=0.31\linewidth]
{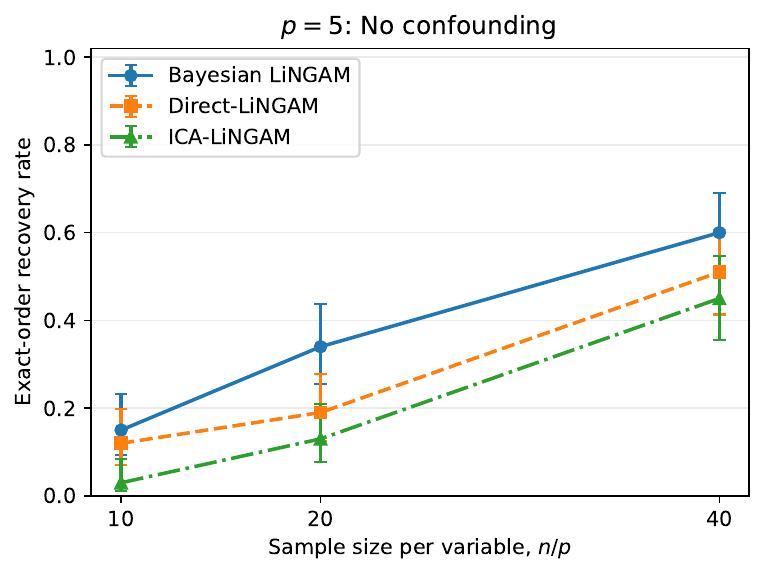}
&
\includegraphics[width=0.31\linewidth]
{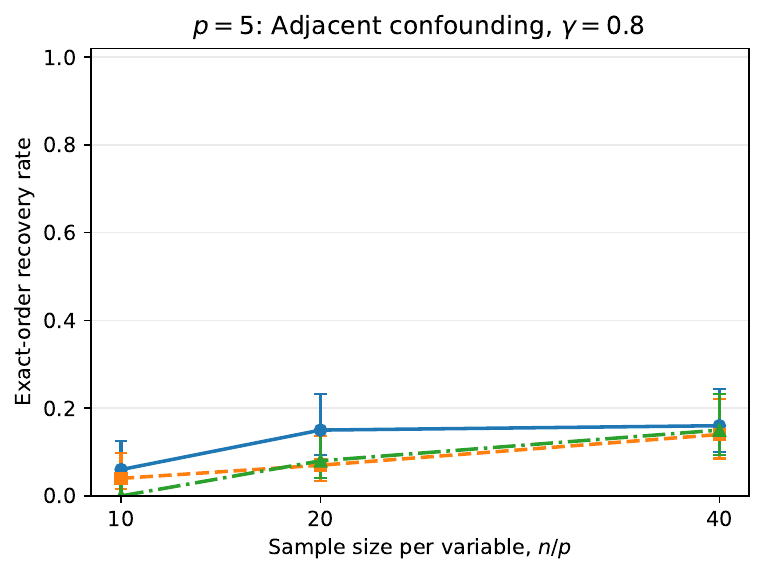}
&
\includegraphics[width=0.31\linewidth]
{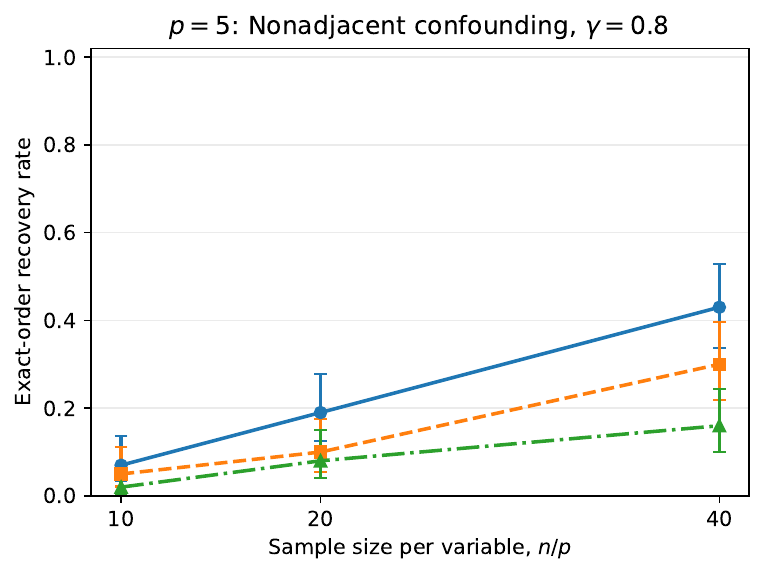}
\\[0.5em]

\multicolumn{3}{c}{\textbf{$p=10$}}
\\[-0.4em]

\includegraphics[width=0.31\linewidth]
{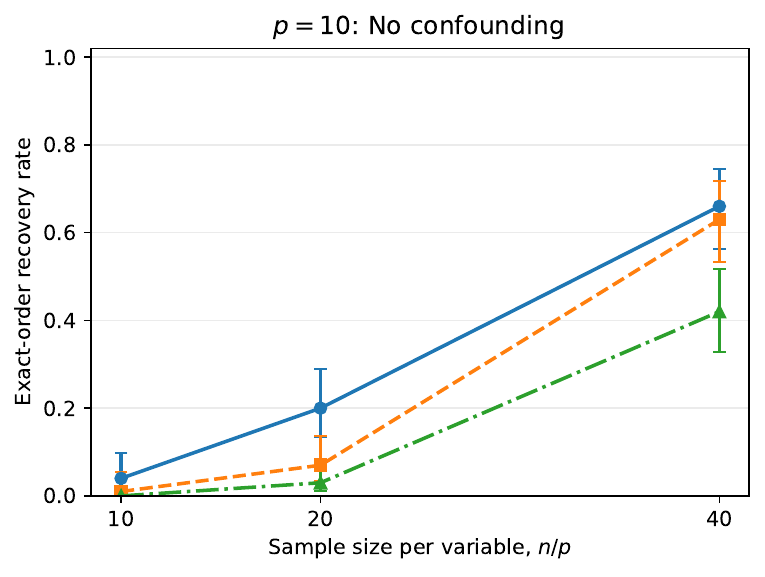}
&
\includegraphics[width=0.31\linewidth]
{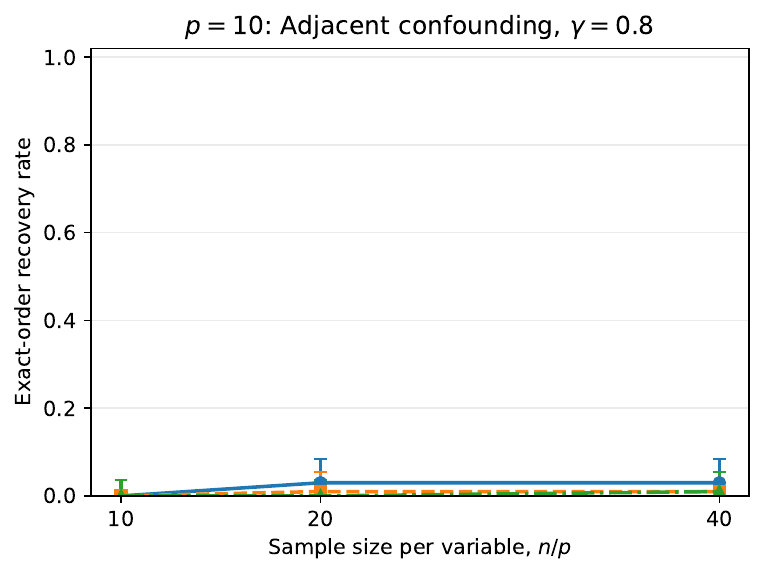}
&
\includegraphics[width=0.31\linewidth]
{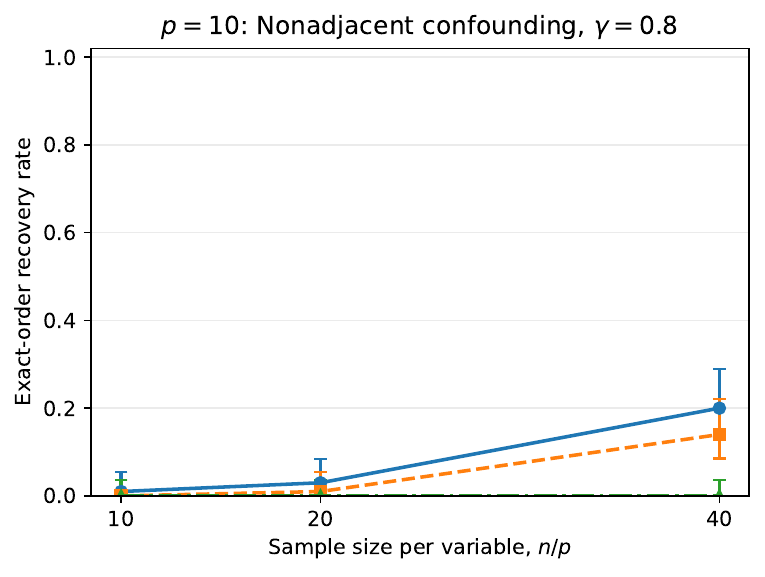}
\\[0.5em]

\multicolumn{3}{c}{\textbf{$p=15$}}
\\[-0.4em]

\includegraphics[width=0.31\linewidth]
{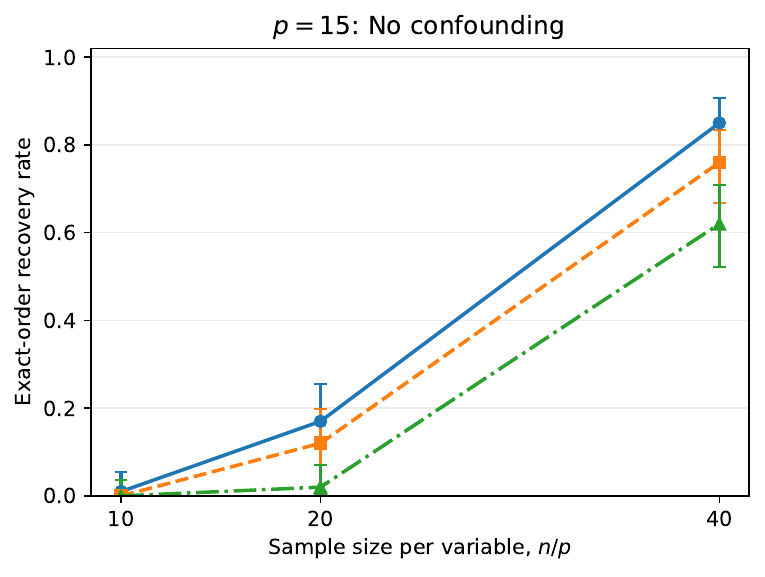}
&
\includegraphics[width=0.31\linewidth]
{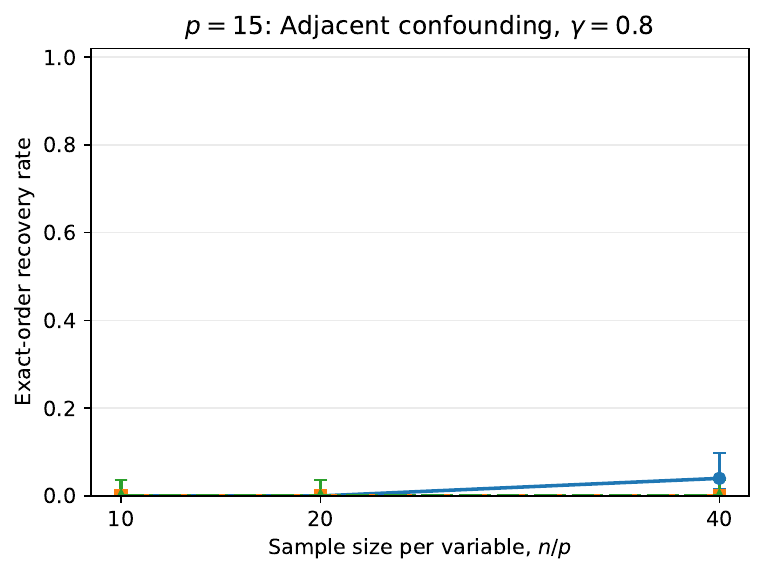}
&
\includegraphics[width=0.31\linewidth]
{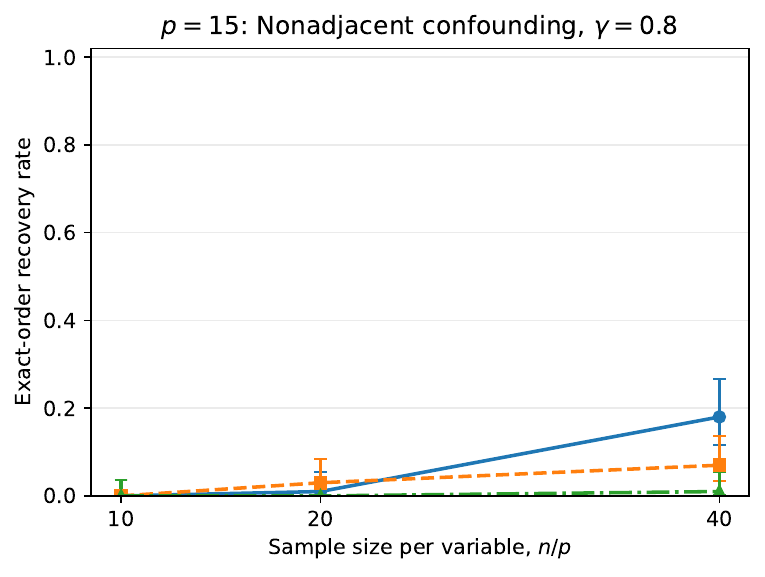}

\end{tabular}

\caption{
Exact-order recovery rates of Bayesian LiNGAM, Direct-LiNGAM, and
ICA-LiNGAM. The upper, middle, and lower rows correspond to $p=5$,
$p=10$, and $p=15$, respectively. The columns show no confounding,
adjacent confounding with $\gamma=0.8$, and nonadjacent confounding
with $\gamma=0.8$. The horizontal axis is the sample size per
variable, $n/p$. Exact recovery requires the entire estimated
permutation to agree with the true causal order. Conditions with zero
exact recoveries are retained and displayed at zero.
}
\label{fig:comparison-exact}
\end{figure}

Figure~\ref{fig:comparison-pairwise} reports the pairwise-order recovery
rates for no confounding and for strong adjacent and nonadjacent
confounding with $\gamma=0.8$. The error bars give 95\% confidence
intervals over the $100$ replications. The results for $\gamma=0.4$
were generally intermediate and are omitted from the figure for
readability. Figure~\ref{fig:comparison-exact} reports the stricter
exact-order recovery rates. Conditions with no exact recoveries are
retained and displayed at zero.

Among the methods evaluated, Bayesian LiNGAM achieved the strongest
overall mean recovery. The main setting-specific findings are as
follows.

\begin{enumerate}
\item
\textbf{Bayesian LiNGAM matched or outperformed the standard LiNGAM
methods in the no-confounding settings.}
DirectLiNGAM and ICA-LiNGAM improved substantially with the sample
size and exhibited strong recovery under the standard LiNGAM
assumptions. At $n/p=40$, the pairwise-order recovery rates of
DirectLiNGAM were approximately $0.84$, $0.91$, and $0.96$ for
$p=5$, $p=10$, and $p=15$, respectively.

Nevertheless, Bayesian LiNGAM attained the highest mean
pairwise-order recovery rates for $p=5$ and $p=10$ at all three
sample sizes and was essentially comparable to DirectLiNGAM for
$p=15$. Its exact-order recovery rates were also higher in most of
the displayed no-confounding conditions. Thus, global comparison of
complete causal orders retained the strong recovery of the standard
LiNGAM methods and produced additional finite-sample improvements,
particularly for $p=5$ and $p=10$.

\item
\textbf{Under adjacent confounding, Bayesian LiNGAM achieved the best
or nearly the best pairwise-order recovery throughout the displayed
conditions.}
For $p=5$ and $p=10$, its mean pairwise-order recovery rate was higher
than that of DirectLiNGAM at all three sample sizes. For $p=15$, the
two methods produced very similar results. 
Thus, although the improvement over DirectLiNGAM was smaller than
under nonadjacent confounding, Bayesian LiNGAM achieved the highest or
nearly the highest mean pairwise-order recovery in the
adjacent-confounding comparison.

Exact-order recovery was difficult for all three methods, particularly
for $p=10$ and $p=15$. In the adjacent design, every confounded pair
is connected by both a directed edge and a latent common cause and
therefore forms a bow. The smaller differences among the methods
suggest that this confounding structure is more difficult for the
complete-order criterion than the nonadjacent structure considered
below.

\item
\textbf{Bayesian LiNGAM showed its clearest and most consistent
advantage under nonadjacent confounding.}
Its mean pairwise-order recovery rate was higher than that of
DirectLiNGAM for every displayed combination of $p$ and $n/p$. At
$n/p=40$, Bayesian LiNGAM attained a pairwise-order recovery rate of
approximately $0.80$ for all three dimensions, whereas that of
DirectLiNGAM was approximately $0.72$--$0.73$.

The exact-order results showed the same advantage. At $n/p=40$, the
exact-order recovery rates of Bayesian LiNGAM were approximately
$0.44$, $0.20$, and $0.18$ for $p=5$, $p=10$, and $p=15$,
respectively, compared with approximately $0.30$, $0.12$, and $0.08$
for DirectLiNGAM. Thus, Bayesian LiNGAM improved both the overall
closeness of the estimated order and the probability of recovering
the complete generating order.

The nonadjacent design is bow-free: directed edges and
latent-confounding relations occur on different variable pairs.
Under non-Gaussianity and suitable genericity conditions,
\citet{WangDrton2023} show that bow-free acyclic path diagrams with
dependent errors can retain higher-order distributional information
for causal identification. 
The present results are consistent with the view that usable ordering
information remains in this setting and that the proposed
complete-order criterion can exploit part of this information more
effectively than the sequence of local scores used by standard
DirectLiNGAM.
The uniform improvement over DirectLiNGAM
across all nine displayed combinations of dimension and sample size is
the main empirical finding of this experiment.
\item
\textbf{The performance of ICA-LiNGAM decreased substantially when
dependence was introduced among the structural disturbances.}
In the no-confounding setting, both its pairwise-order and exact-order
recovery rates increased markedly with the sample size. Under strong
adjacent or nonadjacent confounding, however, its pairwise-order
recovery rate for $p=10$ and $p=15$ remained close to the random-order
level, and its exact-order recovery rate was often zero or nearly
zero.

This decline is expected because the confounded simulations violate
the mutually independent disturbance assumption underlying
ICA-LiNGAM. It therefore illustrates the practical consequence of
applying the standard ICA-LiNGAM criterion when this assumption is
violated, rather than a weakness of ICA-LiNGAM under its intended
model. Bayesian LiNGAM, by contrast, retained a meaningful
complete-order criterion when the disturbances were dependent and
continued to recover substantial causal-order information under both
confounding structures.
\end{enumerate}

\paragraph{Computation time.}

Table~\ref{tab:comparison-time} reports implementation-level computation
times for $p=10$ and $p=15$. For each $(p,n)$, the mean and median were
calculated from 15 sequential fits, comprising five nominal
data-generating conditions and three replications. The proposed method
used the compiled standardized Student-$t_{10}$ shape-score implementation
and ordinary Dijkstra search with \texttt{jobs=1}. The Student-$t$
score kernel was compiled, while OLS was performed by NumPy/LAPACK and
the heap operations were performed in Python. The BLAS/LAPACK thread
pools were restricted to one for this implementation. DirectLiNGAM and
ICA-LiNGAM were executed one fit at a time, although the thread counts of
their underlying numerical libraries were left at the runtime defaults.
Data generation was excluded from all timings; compilation and the
preliminary validation of the compiled score were also excluded from the
Bayesian LiNGAM times.

\begin{table}[!htbp]
\centering
\caption{
Implementation-level computation times in seconds. Each entry reports
the mean, with the median in parentheses, over 15 sequential fits.
}
\label{tab:comparison-time}
\begin{tabular}{ccccc}
\toprule
$p$ & $n$
& Bayesian LiNGAM
& DirectLiNGAM
& ICA-LiNGAM
\\
\midrule
10 & 100 & 0.097 (0.096) & 0.155 (0.155) & 0.244 (0.252) \\
10 & 200 & 0.125 (0.123) & 0.199 (0.195) & 0.275 (0.278) \\
10 & 400 & 0.174 (0.158) & 0.188 (0.170) & 0.148 (0.073) \\
15 & 150 & 4.626 (4.216) & 0.466 (0.466) & 0.342 (0.355) \\
15 & 300 & 6.215 (6.157) & 0.587 (0.522) & 0.366 (0.414) \\
15 & 600 & 9.476 (9.843) & 0.584 (0.551) & 0.146 (0.073) \\
\bottomrule
\end{tabular}
\end{table}

In every timing run, the proposed standardized solver expanded the
complete subset graph: $1024$ nodes and $5120$ edges for $p=10$, and
$32768$ nodes and $245760$ edges for $p=15$. For $p=10$, the recorded
times were of the same order of magnitude for the three implementations.
For $p=15$, complete-subset-graph optimization required substantially
more time than DirectLiNGAM and ICA-LiNGAM, although its mean remained
below ten seconds in all the examined settings.

These values describe the computational scale of the present
implementations rather than a controlled hardware-normalized benchmark.
The timing programs were run separately, their underlying numerical
thread controls were not identical, and different normalizations of the
latent-confounder loading were used in the confounded timing runs. The
values should therefore not be interpreted as paired speed comparisons
on identical confounded data. They are also not a direct empirical test
of Theorem~2: that result concerns asymptotic concentration of the search
under its stated no-confounding conditions, whereas the standardized
implementation used here explored the full subset graph.

\medskip
Overall, the standardized implementation of Bayesian LiNGAM generally
attained the highest or nearly the highest mean recovery under both
criteria in the simulated chain models. It retained competitive recovery
in the no-confounding settings and remained among the best methods under
adjacent confounding. Its clearest advantage occurred under nonadjacent
bow-free confounding, where it yielded higher mean pairwise-order
recovery than standard DirectLiNGAM throughout the displayed conditions
and higher exact-order recovery at $n/p=40$.

These findings provide empirical support for evaluating complete causal
orders through the proposed standardized disturbance-dependence
criterion when the structural disturbances are dependent. The evidence
concerns the combined standardized-score and global-search procedure in
the examined chain models: it does not separately identify the effect of
global optimization, establish the finite-sample performance of the
exact Bayesian marginal-likelihood score, or compare against a method
specifically designed for bow-free confounding. Within these boundaries,
the consistent improvement over the standard baselines in the
nonadjacent design constitutes the main empirical finding of the
experiment.

\paragraph{Code availability.}
The source code and scripts used to produce the numerical results
reported in this paper are publicly available at
\url{https://github.com/prof-joe/bayesian-lingam}.
The exact version used in the experiments (v1.0.0) is permanently
archived on Zenodo at
\url{https://doi.org/10.5281/zenodo.22036228}.

\section{Conclusion}

This study proposed a Bayesian ICA-based method for causal ordering that was motivated by two limitations of existing LiNGAM approaches. First, ICA-LiNGAM may be affected by local solutions and initialization in ICA estimation, whereas DirectLiNGAM constructs a causal ordering through a sequence of local and greedy decisions. In contrast, the proposed method assigns a globally defined score to each candidate causal ordering and searches for the globally optimal ordering by means of a shortest-path algorithm. Second, we provided a principled Bayesian formulation of the ICA criterion. Specifically, the dependence among the disturbance variables was quantified by their multivariate mutual information, or total correlation, which is the Kullback--Leibler divergence between their joint distribution and the product of their marginal distributions. This quantity was estimated from Bayesian marginal likelihoods under the Gaussian copula model. The theoretical results showed that the normalized marginal-likelihood terms converge to the corresponding differential entropies, thereby providing an asymptotic justification for the proposed mutual information criterion.

In the models and simulation settings considered in this study, latent confounding was represented by dependence among the disturbance variables. The proposed method therefore selected a causal ordering by minimizing a Bayesian estimate of this confounding-induced dependence. It should be noted, however, that minimizing multivariate mutual information does not coincide exactly with maximizing the finite-sample recovery rate of the true causal ordering. The former is a distributional criterion, whereas the latter is a discrete evaluation measure of the estimated ordering. Consequently, strict superiority over ICA-LiNGAM and DirectLiNGAM in every finite-sample setting is not necessarily expected.

Nevertheless, the experimental results indicate that the principal objective of the proposed method was largely achieved. In the absence of confounding, ICA-LiNGAM and DirectLiNGAM already exhibited performance close to the attainable ceiling, leaving little room for substantial improvement. Even under such favorable conditions for the existing methods, the proposed method generally achieved comparable performance and sometimes produced further improvements. When confounding was present, the advantage of the proposed method became more evident: it frequently attained higher causal-order recovery rates and smaller pairwise ordering errors than the existing methods. Thus, although ICA-LiNGAM and DirectLiNGAM are already highly effective procedures, the proposed method achieved performance comparable to or better than these methods over a broad range of the experimental conditions, particularly in the more difficult settings involving confounding.

Several issues remain for future research. First, the relationship between minimizing multivariate mutual information and maximizing the probability of recovering the true causal ordering should be investigated more thoroughly. In particular, a finite-sample analysis may clarify why an existing method occasionally achieves a higher recovery rate even when the proposed Bayesian criterion more directly reflects the joint independence of the disturbance variables. Establishing conditions under which minimization of the proposed criterion guarantees recovery of the true ordering would provide a stronger theoretical foundation for the method.

Second, the present study has focused on linear structural equation models with continuous variables. It would be of considerable interest to extend the Bayesian ICA framework to nonlinear causal relationships and to settings involving discrete, binary, or mixed variables. Such extensions would help determine whether the basic principle developed in this study---Bayesian evaluation of the joint independence of disturbances combined with global optimization over causal structures---can serve as a more general framework for causal discovery.

Finally, global optimization using the Bayesian mutual information criterion is computationally more demanding than the existing LiNGAM procedures. Improving the efficiency of the score calculation and developing scalable search strategies for higher-dimensional problems therefore constitute another important direction for future work. We expect that these theoretical, methodological, and computational extensions will further broaden the applicability of Bayesian ICA to causal discovery beyond the linear and continuous setting considered here.

\bibliographystyle{unsrtnat}
\bibliography{ref}

\appendix

\section{Proof of Theorem~\ref{kei1}}
\label{proof-1}
\begin{proof}

Fix a nonempty set $A$ and write
\[
\ell_A^n(\theta_A)
=
\sum_{i=1}^n
\log p_A(z_{i,A}\mid\theta_A)
\qquad 
\text{and}
\qquad
L_A(\theta_A)
=
\mathbb{E}_0\!\left[
\log p_A(Z_A\mid\theta_A)
\right].
\]
Since
$p_{0,A}(z_A)=p_A(z_A\mid\theta_A^0)$,
\[
L_A(\theta_A)-L_A(\theta_A^0)
=
-D_{\mathrm{KL}}\!\left(
 p_{0,A}
 \,\middle\|\,
 p_A(\cdot\mid\theta_A)
\right)
\leq0.
\]
Therefore,
\[
\sup_{\theta_A\in\Theta_A}L_A(\theta_A)
=
L_A(\theta_A^0)
=
\int p_{0,A}(z_A)\log p_{0,A}(z_A)\,dz_A
=
-h(Z_A).
\]

By the compactness, continuity, and integrable-envelope assumptions, the
uniform law of large numbers gives
\[
\sup_{\theta_A\in\Theta_A}
\left|
\frac{1}{n}\ell_A^n(\theta_A)-L_A(\theta_A)
\right|
\ \xrightarrow{p}\ 0.
\]
We first prove the upper bound. For every $\varepsilon>0$, with
probability tending to one,
\[
\frac{1}{n}\ell_A^n(\theta_A)
\leq
-h(Z_A)+\varepsilon
\]
for all $\theta_A\in\Theta_A$. Since $\pi_A$ is a probability measure,
\[
g_A^n
\leq
\exp\left\{
n\bigl[-h(Z_A)+\varepsilon\bigr]
\right\}.
\]
Hence
\begin{equation}\label{eq41}
\limsup_{n\to\infty}
\frac{1}{n}\log g_A^n
\leq
-h(Z_A)
\end{equation}
in probability.

We next prove the lower bound. By the continuity of $L_A$ at
$\theta_A^0$, for every $\varepsilon>0$ there exists a neighborhood
$N_\varepsilon\subset\Theta_A$ of $\theta_A^0$ such that
\[
L_A(\theta_A)
\geq
-h(Z_A)-\varepsilon
\]
for all $\theta_A\in N_\varepsilon$. By assumption,
$\pi_A(N_\varepsilon)>0$. The uniform law of large numbers implies that,
with probability tending to one,
\[
\frac{1}{n}\ell_A^n(\theta_A)
\geq
-h(Z_A)-2\varepsilon
\]
for all $\theta_A\in N_\varepsilon$. Consequently,
\[
g_A^n
\geq
\pi_A(N_\varepsilon)
\exp\left\{
n\bigl[-h(Z_A)-2\varepsilon\bigr]
\right\}.
\]
Since $\pi_A(N_\varepsilon)$ does not depend on $n$,
\begin{equation}\label{eq42}
\liminf_{n\to\infty}
\frac{1}{n}\log g_A^n
\geq
-h(Z_A)-2\varepsilon
\end{equation}
in probability. Since $\varepsilon>0$ is arbitrary,
from \eqref{eq41} and \eqref{eq42}, we have
\[
\frac{1}{n}\log g_A^n
\ \xrightarrow{p}\
-h(Z_A),
\]
which proves
\[
-\frac{1}{n}\log g_A^n
\ \xrightarrow{p}\
h(Z_A).
\]

Let $\widetilde g_A^n$ denote the Laplace approximation in
Section~\ref{subsec:gaussian-copula-marginal}. Under the corresponding
regularity conditions,
\[
\log g_A^n
=
\log\widetilde g_A^n+O_p(n^{-1}).
\]
Therefore,
\[
-\frac{1}{n}\log\widetilde g_A^n
=
-\frac{1}{n}\log g_A^n+O_p(n^{-2})
\ \xrightarrow{p}\
h(Z_A).
\]
\end{proof}

\section{Proof of Theorem~\ref{teiri2}}
\label{proof-2}

Lemma~\ref{hojo:DS} states that a component appearing with nonzero
coefficients in two independent linear forms must be Gaussian.
In the proof below, the first linear form is the competing variable $X_k'$,
and the second is one coordinate $R_r^{(j,k)}$ of the corresponding
residual vector. Thus, the lemma first excludes every non-Gaussian
component shared by $X_k'$ and $R_r^{(j,k)}$. The condition in
Theorem~\ref{teiri2} that at least $p-1$ residual variables are
non-Gaussian, together with the finite positive variance condition,
then excludes the remaining possible shared Gaussian component. See,
for example, \citet{kagan1973characterization}.

\begin{hojo}[Darmois--Skitovich]\label{hojo:DS}
Let $U_1,\ldots,U_q$ be mutually independent nondegenerate random
variables. If
\[
\sum_{\ell=1}^q a_\ell U_\ell,
\qquad
\text{and}
\qquad
\sum_{\ell=1}^q b_\ell U_\ell
\]
are independent, then every $U_\ell$ satisfying
\[
a_\ell b_\ell\neq0
\]
is Gaussian.
\end{hojo}

\begin{proof}

Since no confounding exists, the residual variables
$Z_1^{\tau^*},\ldots,Z_p^{\tau^*}$
along the order $\tau^*$ are mutually independent. Hence,
\[
K_{\tau^*}=0.
\]
The finite positive variance condition 
assumed in Theorem~\ref{teiri2}
ensures that all the population regression coefficients used below are
well defined.

Consider a node on the path corresponding to $\tau^*$ after the first
$j-1$ variables have been selected. After removing the effects of
these variables, write the remaining variables as
\[
{X}^{(j)}
=
\left[
\begin{array}{c}
X_j'\\
\vdots\\
X_p'
\end{array}
\right]
=
A^{(j)}
{Z}^{(j)}
\quad\text{and}\quad
{Z}^{(j)}
=
\left[
\begin{array}{c}
Z_j^{\tau^*}\\
\vdots\\
Z_p^{\tau^*}
\end{array}
\right],
\]
where $A^{(j)}$ is lower triangular with unit diagonal. In particular,
\[
X_j'=Z_j^{\tau^*}.
\]
For every $r>j$, we have
\[
X_r'
=
a_{rj}^{(j)}Z_j^{\tau^*}
+
\sum_{\ell=j+1}^{r}
a_{r\ell}^{(j)}Z_\ell^{\tau^*}.
\]
Since $Z_j^{\tau^*}$ is independent of
$Z_{j+1}^{\tau^*},\ldots,Z_p^{\tau^*}$, the population regression
coefficient of $X_r'$ on $X_j'$ is $a_{rj}^{(j)}$, and the corresponding
residual is
\[
\sum_{\ell=j+1}^{r}
a_{r\ell}^{(j)}Z_\ell^{\tau^*},
\]
which is independent of $X_j'$. Thus, the edge corresponding to the
true next variable has population length zero.

Fix $k>j$ and consider the competing edge that selects $X_k'$ next.
Let
\[
{R}^{(j,k)}
=
\left(
R_r^{(j,k)}
\right)_{r\in\{j,\ldots,p\}\setminus\{k\}},
\qquad
R_r^{(j,k)}
=
X_r'
-
\frac{\operatorname{Cov}(X_r',X_k')}
     {\operatorname{Var}(X_k')}
X_k'.
\]
Suppose, to the contrary, that this competing edge also has zero
population length:
\begin{equation}\label{eq31}
I\left(
X_k';{R}^{(j,k)}
\right)=0.
\end{equation}
Then,
\[
X_k'\ci {R}^{(j,k)},
\]
and hence
\[
X_k'\ci R_r^{(j,k)}
\qquad
\text{for every }r\in\{j,\ldots,p\}\setminus\{k\}.
\]
To obtain the desired contradiction, we first show that
\eqref{eq31} implies
\begin{equation}\label{eq32}
X_k'=Z_k^{\tau^*}.
\end{equation}
Note that the transformation
\[
{\mathbb R}^{p-j+1}\ni {X}^{(j)}
\longmapsto
\left[
\begin{array}{c}
X_k'\\
{R}^{(j,k)}
\end{array}
\right]
\in{\mathbb R}^{p-j+1}
\]
is nonsingular, because each $X_r'$ can be recovered from $X_k'$ and
$R_r^{(j,k)}$. Hence,
there exists a
nonsingular matrix
$M^{(j,k)}=
\left[
\begin{array}{c}
m_1^\top\\
\vdots\\
m_q^\top
\end{array}
\right]
\in{\mathbb R}^{q\times q}$ with $q:=p-j+1$
such that 
\[
\left[
\begin{array}{c}
X_k'\\
{R}^{(j,k)}
\end{array}
\right]
=
M^{(j,k)}{Z}^{(j)}.
\]
We index the columns of $M^{(j,k)}$ by
$\ell=j,\ldots,p$, and let $m_{s\ell}$ denote the entry in row $s$
and the column corresponding to $Z_\ell^{\tau^*}$. Thus,
\[
X_k'
=
m_1^\top {Z}^{(j)},
\]
and each coordinate of ${R}^{(j,k)}$ is of the form
\[
m_s^\top {Z}^{(j)},
\qquad
s=2,\ldots,q.
\]
Fix $s\in\{2,\ldots,q\}$. Since
\[
m_1^\top {Z}^{(j)}
\ci
m_s^\top {Z}^{(j)},
\]
Lemma~\ref{hojo:DS} implies that, for every non-Gaussian component
$Z_\ell^{\tau^*}$,
\[
m_{1\ell}m_{s\ell}=0.
\]
Thus, no non-Gaussian component can appear with nonzero coefficients
in both $X_k'$ and the $s$th coordinate of ${R}^{(j,k)}$.
By the non-Gaussianity condition in Theorem~\ref{teiri2},
there is at most one Gaussian component among the coordinates of
${Z}^{(j)}$. If such a component exists, denote its index by $g$.
Since the two linear forms are independent and have finite variances,
\[
0
=
\operatorname{Cov}
\left(
m_1^\top {Z}^{(j)},
m_s^\top {Z}^{(j)}
\right).
\]
The components of ${Z}^{(j)}$ are mutually independent, and
Lemma~\ref{hojo:DS} has already shown that
$m_{1\ell}m_{s\ell}=0$ for every non-Gaussian index $\ell$. Hence,
\[
0
=
m_{1g}m_{sg}
\operatorname{Var}\!\left(Z_g^{\tau^*}\right).
\]
Since
\[
0<
\operatorname{Var}\!\left(Z_g^{\tau^*}\right)
<\infty,
\]
we obtain
\[
m_{1g}m_{sg}=0.
\]
Therefore, whether or not a Gaussian component exists,
\[
m_{1\ell}m_{s\ell}=0
\qquad
\text{for every }\ell=j,\ldots,p
\quad\text{and}\quad
s=2,\ldots,q.
\]
Let
\[
\mathcal J
=
\left\{
\ell\in\{j,\ldots,p\}:m_{1\ell}\neq0
\right\}.
\]
For every $\ell\in\mathcal J$, the $\ell$th column of $M^{(j,k)}$
is zero except for its first entry. If $\mathcal J$ contained two or
more elements, the corresponding columns of $M^{(j,k)}$ would all be
nonzero scalar multiples of the first coordinate vector. They would
therefore be linearly dependent, contradicting the nonsingularity of
$M^{(j,k)}$. Consequently,
\[
|\mathcal J|=1.
\]
It follows that, for some $\ell\in\{j,\ldots,p\}$ and some nonzero
constant $c$,
\[
X_k'=cZ_\ell^{\tau^*}.
\]
On the other hand, since $A^{(j)}$ is lower triangular with unit
diagonal,
\[
X_k'
=
\sum_{r=j}^{k-1}
a_{kr}^{(j)}Z_r^{\tau^*}
+
Z_k^{\tau^*}.
\]
Consequently,
\[
\ell=k,\qquad c=1,
\]
and hence we obtain \eqref{eq32}.

Let $\widetilde{\tau}$ be the order obtained by moving
$\tau_k^*$ to position $j$ while preserving the relative order of
all the other variables:
\[
\widetilde{\tau}
=
\left(
\tau_1^*,\ldots,\tau_{j-1}^*,
\tau_k^*,
\tau_j^*,\tau_{j+1}^*,\ldots,\tau_{k-1}^*,
\tau_{k+1}^*,\ldots,\tau_p^*
\right).
\]
Since $X_k'=Z_k^{\tau^*}$, the variable $X_k'$ is independent of
$X_j',\ldots,X_{k-1}'$. Moving it forward therefore does not violate
the triangular structural equations. The relative order of all the
other variables is unchanged, and the residual variables along
$\widetilde{\tau}$ are
\[
Z_1^{\tau^*},\ldots,Z_{j-1}^{\tau^*},
Z_k^{\tau^*},
Z_j^{\tau^*},\ldots,Z_{k-1}^{\tau^*},
Z_{k+1}^{\tau^*},\ldots,Z_p^{\tau^*}.
\]
They are mutually independent, and hence
\[
K_{\widetilde{\tau}}=0.
\]
Since $\widetilde{\tau}\neq\tau^*$, this contradicts the uniqueness
of $\tau^*$ in Assumption~\ref{katei1}. Therefore, every competing
edge at a node on the path corresponding to $\tau^*$ has strictly
positive population length.

There are only finitely many such competing edges. Hence,
\[
\Delta
:=
\min_{1\leq j<k\leq p}
I\left(
X_k';{R}^{(j,k)}
\right)
>0.
\]
By the consistency of $I_n(\cdot;\cdot)$ and the finiteness of the
set of edges, after replacing negative estimates by zero as in
Section~3.2, with probability tending to one, every estimated edge
length on the path corresponding to $\tau^*$ is smaller than
$\Delta/(4p)$, whereas every competing edge length leaving this path
is larger than $\Delta/2$.

On this event, the estimated cumulative distance of every node on the
path corresponding to $\tau^*$ is smaller than
\[
p\frac{\Delta}{4p}
=
\frac{\Delta}{4}.
\]
In contrast, every path leaving the true path contains at least one
competing edge. Since all the edge lengths are nonnegative, the
cumulative distance of every such path is larger than
$\Delta/2$. It follows inductively that the OPEN/CLOSED search always
selects the next node on the path corresponding to $\tau^*$ and never
expands a node outside that path.

The algorithm evaluates $p$ outgoing edges at the initial node,
$p-1$ edges at the next node, and so on, until two variables remain.
Therefore, the total number of mutual information evaluations is
\[
p+(p-1)+\cdots+2
=
\frac{p(p+1)}{2}-1.
\]
DirectLiNGAM likewise evaluates every remaining candidate variable at
each stage, and hence performs the same number of candidate-variable
evaluations.

\end{proof}

\section{Standardized Implementation of the Proposed Score}
\label{app:standardized-score}

The large comparison used a computationally efficient standardized
implementation of the proposed ordering criterion. For an order
$\tau=(\tau_1,\ldots,\tau_p)$, let $\widehat e_j^\tau$ be the OLS
residual obtained by regressing $X_{\tau_j}$ on its predecessors, and
put
\[
(\widehat s_j^\tau)^2
=\frac{1}{n}\|\widehat e_j^\tau\|^2.
\]
Each residual vector was centered and divided by
$\widehat s_j^\tau$, and the following standardized Student-$t$
shape score was evaluated with $\nu=10$:
\[
\widehat C_{\mathrm{std}}(\tau)
=
-\frac{1}{n}
\sum_{j=1}^p
\sum_{i=1}^n
\log q_{10}
\left(
\frac{\widehat e_{ij}^\tau}
     {\widehat s_j^\tau}
\right),
\]
where $q_{10}$ denotes the unit-variance Student-$t_{10}$ density.

This standardization does not discard an order-dependent residual-scale
term at the plug-in likelihood level. Let
$\widehat\Sigma=n^{-1}X_c^{\mathsf T}X_c$ be the empirical covariance
matrix of the centered observations. The Schur-complement identity gives
\[
\prod_{j=1}^p(\widehat s_j^\tau)^2
=\det\widehat\Sigma,
\qquad
\sum_{j=1}^p\log\widehat s_j^\tau
=\frac{1}{2}\log\det\widehat\Sigma,
\]
which is common to all complete orders. Thus, standardizing the
residuals preserves the comparison of complete orders for the plug-in
location-scale likelihood. The identity is exact for ordinary least
squares. The implementation used a ridge of $10^{-10}$ and a residual
scale floor of $10^{-8}$ only for numerical stability, so the identity
holds up to this numerical regularization.

The standardized implementation retains the leading likelihood term but
does not include the prior and observed-information terms in the
finite-sample Laplace expansion. After normalization by $n$, its
order-dependent omitted terms are of order $n^{-1}$ under the regularity
conditions of Section~3. It should therefore be regarded as a
computational surrogate for, rather than the exact finite-sample value
of, the Bayesian marginal-likelihood score. Under the correctly
specified unconfounded Student-$t_{10}$ model, it has the same separated
population minimizer as the corresponding Bayesian score. No such
equivalence is asserted for the misspecified confounded conditions.
Because the standardized edge costs are positive, their sum was
minimized exactly by ordinary Dijkstra search; the dynamic-programming
fallback described in Section~4.1 was not used for this implementation.

We assessed the effect of this computational approximation by also
running a nonstandardized location-scale Laplace implementation on all
$1{,}500$ matched data sets for $p=10$, comprising three sample sizes,
five data-generating conditions, and $100$ replications. This version
estimated the location and scale of each unstandardized residual vector
and retained the observed-information correction, with the prior term
held constant. The two implementations selected exactly the same
complete order in $75.5\%$ of the data sets. Their overall mean
pairwise-order recovery rates were $0.7182$ for the standardized version
and $0.7173$ for the nonstandardized version, and their exact-order
recovery rates were $0.1547$ and $0.1520$, respectively. Across the 15
conditions, the absolute difference between their mean pairwise-order
recovery rates was at most $0.014$, and that between their exact-order
recovery rates was at most $0.03$. In this head-to-head implementation,
the corresponding mean computation times were $0.813$ and $22.143$
seconds, so that the standardized version was approximately 27 times
faster. These results motivated its use in the main comparison: it
substantially reduced computation time without materially changing the
recovery conclusions in the matched $p=10$ sensitivity experiment.

\end{document}